\documentclass[a4paper,11pt]{article}
\usepackage{jcappub} 

\usepackage{cleveref}
\usepackage{amsmath}
\usepackage{tabularx}
\usepackage{booktabs}
\usepackage{subcaption}

\newcommand{\Prob}{\text{P}}           
\newcommand{\posterior}{\mathcal{P}}   
\newcommand{\likelihood}{\mathcal{L}}  
\newcommand{\prior}{\pi}               
\newcommand{\evidence}{\mathcal{Z}}    
\newcommand{\data}{D}        
\newcommand{\model}{\mathcal{M}}       
\newcommand{\params}{\theta}           
\newcommand{\paramsM}{\theta_\mathcal{M}} 
\newcommand{\KL}{\mathcal{D}_{\text{KL}}}    
\newcommand{\vprior}{V_\pi}          
\newcommand{\shannon}{\mathcal{I}}  
\newcommand{\vposterior}{V_\mathcal{P}} 

\arxivnumber{2511.04661} 
\title{\boldmath \texttt{unimpeded}: A Public Grid of Nested Sampling Chains for Cosmological Model Comparison and Tension Analysis}

\author[1,2]{Dily Duan Yi Ong\note{Corresponding author.}}
\author[1,2]{and Will Handley}
\affiliation[1]{Kavli Institute for Cosmology, University of Cambridge,\\Madingley Road, Cambridge, CB3 0HA, U.K.}
\affiliation[2]{Cavendish Laboratory, University of Cambridge,\\J.J. Thomson Avenue, Cambridge, CB3 0HE, U.K.}

\emailAdd{dlo26@cam.ac.uk}

\abstract{Bayesian inference is central to modern cosmology, yet comprehensive model comparison and tension quantification remain computationally prohibitive for many researchers. To address this, we release \texttt{unimpeded}, a publicly available Python library and data repository providing pre-computed nested sampling and MCMC chains. We apply this resource to conduct a systematic analysis across a grid of eight cosmological models, including $\Lambda$CDM and seven extensions, and 39 datasets, including individual probes and their pairwise combinations. Our model comparison reveals that whilst individual datasets show varied preferences for model extensions, the base $\Lambda$CDM model is most frequently preferred in combined analyses, with the general trend suggesting that evidence for new physics is diluted when probes are combined. Using five complementary statistics, we quantify tensions, finding the most significant to be between DES and Planck ($\sigma=3.57{\scriptstyle\pm0.10}$) and SH0ES and Planck ($\sigma=3.27{\scriptstyle\pm0.10}$) within $\Omega_k\Lambda$CDM. We characterise the $S_8$ tension as high-dimensional ($d_G=4.87{\scriptstyle\pm0.79}$) and partially resolvable in certain extended models, whereas the Hubble tension is low-dimensional and persists across the model space. Caution should be exercised when combining datasets in tension. The \texttt{unimpeded} data products, hosted on Zenodo, provide a powerful resource for reproducible cosmological analysis and underscore the robustness of the $\Lambda$CDM model against this comprehensive benchmark compilation.
}

\begin{document}
\maketitle
\flushbottom

\section{Introduction}
\label{sec:introduction}
Bayesian methods of inference are widely used in modern cosmology in parameter estimation, model comparison and tension quantification. Parameter estimation refers to the process of determining the values of cosmological parameters, which describe the properties of a model using observed data. Model comparison refers to the evaluation and selection between different cosmological models and tension quantification is the measurement and study of discrepancies between different observed datasets, which are predicted to be in agreement theoretically by a cosmological model. The last two of these have gained more prominence in recent times due to disparities that have emerged within the context of the concordance model regarding the estimated value of the Hubble constant $H_{0}$~\cite{Verde2019NatAs} using Cosmic Microwave Background (CMB) and Supernovae data (commonly referred to as the Hubble tension), the clustering $\sigma_{8}$~\cite{Joudaki2017MNRAS} using CMB and weak lensing, and the curvature $\Omega_{K}$~\cite{Handley2021PRD,DiValentino2020NatAs} using CMB and lensing/BAO, and between CMB datasets. These tensions, may excitingly, point towards physics beyond the standard $\Lambda$ Cold Dark Matter ($\Lambda$CDM) concordance model. 

Parameter estimation has commonly been performed using the Markov chain Monte Carlo (MCMC) methods, which are effective for exploring the posterior distributions of model parameters given a set of data and a model. The Planck Legacy Archive (PLA)~\cite{Planck2018params} has been an invaluable community resource, providing MCMC chains for a grid of models and datasets, primarily facilitating parameter estimation.  However, MCMC methods are not suitable for calculating Bayesian evidence, which is essential for model comparison and tension quantification. Nested sampling~\cite{Skilling2006,Lemos2021,Handley2019,Lemos2020} has emerged as a powerful alternative, specifically tailored for model comparison and tension quantification. It is a Monte Carlo sampling technique used to efficiently compute the evidence and concurrently generate samples from the posterior distribution as a by-product, and hence, enabling parameter estimation without extra expense. However, the computational cost of nested sampling is still significant, especially when considering the large parameter spaces and complex likelihoods involved in modern cosmological analyses.

This paper introduces \texttt{unimpeded}~\cite{2025arXiv251105470O}\footnote{The \texttt{unimpeded} library and its source code are available at \url{https://github.com/handley-lab/unimpeded}.}, a publicly available pip-installable Python library and associated data repository. The primary aim of \texttt{unimpeded} is to provide an analogous grid to the PLA but utilising nested sampling chains, thereby enabling robust model comparison and tension quantification alongside parameter estimation. This initiative directly supports the goals of our DiRAC-funded projects (DP192 and 264), which seek a systematic examination of model-dataset combinations to uncover patterns that might illuminate the path towards resolving current cosmological puzzles or identifying a successor to $\Lambda$CDM. In this study, we focus our analysis on the standard $\Lambda$CDM model and seven of its most widely studied extensions, including both single-parameter and two-parameter variants. We acknowledge that this choice is conservative and represents a specific, well-defined subset of the vast landscape of possible cosmological models. Our primary goal is to assess how evidence for these canonical extensions evolves with dataset combination, rather than to perform an exhaustive search for new physics. Therefore, our conclusions regarding tensions and model preference should be interpreted as being conditional on this specific, and intentionally restricted, model space. While the current release focuses on these baseline models, the deliberate use of wide priors positions the chains for efficient extension to alternative models via importance sampling techniques, which will be integrated into future versions of \texttt{unimpeded} to allow rapid testing of additional theoretical scenarios without requiring full re-sampling. The \texttt{unimpeded} grid is designed to incorporate a broad variety of modern datasets and expanding as new data and models become relevant. All associated data products, including nested sampling and MCMC chains, are made publicly and permanently available on Zenodo\footnote{\url{https://zenodo.org/}}. Newer datasets, including DESI DR2, Pantheon+, Union3, and DES Y5, continue to be analysed and their nested sampling chains are made available publicly in our ongoing work~\cite{2025arXiv251110631O, 2026arXiv260305472O}.

This paper is structured as follows. In Section~\ref{sec:theory}, we review the theoretical foundations of Bayesian inference and the three pillars of Bayesian cosmological analysis, namely parameter estimation, model comparison and tension quantification. We also discuss the concordance $\Lambda$CDM model and other models. We then detail our specific methodological approach with nested sampling in Section~\ref{sec:methodology}. The \texttt{unimpeded} library and its core functionalities are introduced in Section~\ref{sec:unimpeded_action}. We apply this framework to cosmological data, presenting our main findings in Section~\ref{sec:results}, presenting wide grids of model comparison and tension statistics.

\section{Theory}
\label{sec:theory}
\subsection{Nomenclature}
\label{ssec:nomenclature}
The Universe provides a single laboratory for physics, but with experimental settings we can only observe, not control. Our goal is to utilises observables to rigorously test predictive cosmological models, seeking either to quantify how likely they are to be true given the data, and to improve their parameter constraints. Bayesian inference provides a framework for systematically updating our beliefs about model and their parameters in light of new data. 

A predictive model $\model$ contains a set of variable parameters $\params$, with some observed dataset $\data$. $\data$ is typically a collection of measurements or observations, such as the Cosmic Microwave Background (CMB), baryon acoustic oscillations (BAO), supernovae data, weak lensing data and gravitational waves. Cosmological models $\model$ are theoretical frameworks that describe the physical properties and evolution of the Universe, typically expressed by a metric with a set of cosmological parameters $\paramsM$. $\paramsM$ has the subscript $\model$ to indicate that the parameters are specific to the model $\model$. For example, the concordance $\Lambda$CDM model has 6 parameters: the Hubble constant $H_0$, the baryon density $\Omega_b h^2$, the cold dark matter density $\Omega_c h^2$, the scalar spectral index $n_s$, the amplitude of primordial scalar perturbations $A_s$ and the reionisation optical depth $\tau_{\mathrm{reio}}$. When only one model is considered, we can drop the subscript $\model$ and write $\params$ instead of $\paramsM$.

\subsubsection{Bayesian Inference}

Before analysing any data, we can express our beliefs about the parameters $\params$ for a specific model $\model$, termed the prior,
\begin{equation}
    \Prob(\params | \model) \equiv \prior(\params) .
\end{equation}
Common choices for the prior include uniform or log-uniform distributions over a range of theoretically and physically allowed values or Gaussian distributions centred around expected values. After considering the observed data $\data$, we can update our beliefs about the parameters $\params$, termed the posterior,
\begin{equation}
    \Prob(\params | \data, \model) \equiv {\posterior}(\params).
\end{equation}
Both the prior and posterior are probability density functions (PDFs), which integrate to 1 over all possible values of $\params$.

The likelihood function describes how probable the observed data $\data$ is given a specific set of cosmological parameter values $\params$ and a specific model $\model$,
\begin{equation}
    \Prob(\data | \params, \model) \equiv \likelihood(\params).
\end{equation}
While $\Prob(\data | \params, \model)$ treats $\data$ as the variable and $\params$ and $\model$ as fixed parameters, $\likelihood(\params)$ treats $\params$ as the variable. This equivalence represents a shift from prediction to inference, which is fundamental in cosmology as we only have one universe to observe. We cannot create new, independent universes to measure, and therefore only have one $\data$. $\Prob(\data | \params, \model)$ is therefore repurposed into a function of $\params$, which we can test with a wide range of $\params$ values and find the maximised $\likelihood(\params)$ that explains our single cosmic observation. Likelihood is not a PDF like prior and posterior, and it does not necessarily integrate to 1 over $\params$ space.

The evidence, or the marginal likelihood~\cite{2008ConPh..49...71T}, is the probability of observing $\data$ given $M$, derived from the likelihood by integrating over all parameters and weighted by the prior,
\begin{equation}
    \Prob(\data | \model) \equiv \evidence = \int \Prob(\data | \params, \model) \Prob(\params | \model) d\params.
\end{equation}
Dropping the model dependence, we have:
\begin{equation}
    \evidence = \int \likelihood(\params) \prior(\params) d\params.
\end{equation}
It can be intuitively understood as a ``prior-weighted average likelihood''. Mathematically, it is the normalising constant that ensures the posterior integrates to unity. The evidence is usually ignored during parameter estimation (see~\Cref{ssec:param_estimation}) but plays a crucial role in model comparison (see~\Cref{ssec:model_comparison}) and tension quantification (see~\Cref{ssec:tension_quant_theory}).

\subsubsection{Kullback-Leibler Divergence}
\label{sssec:kl_divergence}

The Kullback-Leibler (KL) divergence, $\KL$, quantifies the information gain, or compression, between the prior distribution $\prior(\params)$ and the posterior distribution $\posterior(\params)$~\cite{kullback1951information}. It has been widely used by cosmologists~\cite{2014PhRvD..90b3533S,2019JCAP...01..011N,2004PhRvL..92n1302H,2013PDU.....2..166V,2016PhRvD..93j3507S,2016JCAP...05..034G,2016arXiv160606273R,2016MNRAS.455.2461H,2016MNRAS.463.1416G,2017NatAs...1..627Z,2017JCAP...10..045N} and is defined as the average of the Shannon Information, $\shannon(\params)$, over the posterior:
\begin{equation}
    \shannon(\params) = \log\frac{\posterior(\params)}{\prior(\params)},
    \label{eq:shannon_info}
\end{equation}
\begin{equation}
    \KL = \int \posterior(\params) \log\frac{\posterior(\params)}{\prior(\params)}\,d\params = \left\langle \log\frac{\mathcal{P}}{\pi}\right\rangle_\mathcal{P} = \left\langle \shannon \right\rangle_{\posterior} \approx \log\left(\frac{V_{\pi}}{V_{P}}\right).
    \label{eq:kl_divergence_def}
\end{equation}
A higher $\KL$ indicates a larger information gain when moving from the prior to the posterior and is consequently a useful measure of the constraining power of the data. $\KL$ can be understood as approximately the logarithm of the ratio of the prior volume, $V_\pi$, to the posterior volume, $V_\mathcal{P}$. This relationship is exact in the case of uniform (``top-hat'') prior and posterior distributions, but remains highly accurate when a broad prior is used, where the prior is 'locally flat" around the posterior peak.

$\KL$ is a strong function of the prior, and it inherits the property of being additive for independent parameters from the Shannon Information. A key practical consideration is that its calculation requires a properly normalised posterior distribution, which in turn requires knowledge of the Bayesian evidence, $\evidence$. Consequently, this quantity is not attainable with common MCMC sampling techniques, which typically generate samples from an unnormalised posterior. To compute it, more computationally intensive algorithms such as nested sampling are necessary.

The KL divergence can be related directly to the Bayesian evidence via the expression~\cite{Hergt2021Bayesian}:
\begin{equation}
	\log \evidence = \langle \log \likelihood \rangle_{\posterior} - \KL,
\label{eq:logZ_KL}
\end{equation}
where $\langle \log \likelihood \rangle_{\posterior}$ is the posterior average of the log-likelihood. This relation, sometimes referred to as the Occam's Razor equation~\cite{Handley_dimensionality_2019}, illustrates how the evidence naturally implements Occam's Razor by penalising unnecessary model complexity. As discussed further in~\Cref{ssec:model_comparison}, the penalty factor between competing cosmological models can be approximated using the difference in their respective $\KL$. Using the nested sampling chains from \texttt{unimpeded}, we compute $\KL$ for every model-dataset combination in our grid, enabling a systematic comparison of the constraining power of different datasets and models. These results are presented in~\Cref{ssec:constraining_power}.

\subsection{Parameter Estimation}
\label{ssec:param_estimation}
The goal of parameter estimation is to determine the posterior probability distribution of the parameters, $\Prob(\params | \data, \model)$. Combining the prior, likelihood and evidence from~\Cref{ssec:nomenclature}, $\Prob(\params | \data, \model)$ is stated by the Bayes' theorem as:
\begin{align}
    \Prob(\params | \data, \model) &= \frac{\Prob(\data | \params, \model) \Prob(\params | \model)}{\Prob(\data | \model)}, \\
    \posterior(\params) &= \frac{\likelihood(\params) \times \prior(\params)}{\evidence}.
\end{align}
It describes how our initial beliefs $\prior(\params)$ about $\params$ are updated in light of the observed data $\data$ under the assumed model $\model$. MCMC algorithms are typically used for exploring the posterior, particularly in high-dimensional parameter spaces~\cite{2002PhRvD..66j3511L}. However, the evidence cannot be obtained due to technical reasons, so MCMC methods give unnormalised posterior $\Prob(\params | \data, \model) \propto \likelihood(\params) \times \prior(\params)$. However, specialised algorithms like nested sampling, as implemented in codes such as \texttt{PolyChord}~\cite{Handley2015PolychordI, Handley2015PolychordII}, can also efficiently generate posterior samples while simultaneously calculating the evidence. The \texttt{unimpeded} library provides access to both nested sampling and MCMC chains for all model-dataset combinations, enabling comprehensive parameter estimation analysis. We present parameter constraints for key cosmological parameters across our grid in~\Cref{ssec:parameter_estimation}.

\subsection{Model Comparison}
\label{ssec:model_comparison}
Model comparison addresses the question of how much the data $\data$ support each competing models \{$\model_1,\model_2,\cdots$\}, where each model $\model_i$ has its own set of parameters $\theta_{\model_i}$. The goal is to compute the posterior $\Prob(\model_i | \data)$ of a model $\model_i$ is true given the data $\data$, which can be used to rank and select models. From the Bayes' theorem, we have:
\begin{align}
  \Prob(\model_i|\data) &= \frac{\Prob(\data|\model_i) \Prob(\model_i)}{\Prob(\data)}, \\
  &= \frac{\evidence_i\prior_i}{\displaystyle\sum_j \evidence_j\prior_j}.
\label{eq:model_posterior}
\end{align}
Evidence is the inner product of the likelihood function and the prior function over the parameter space, it can also be viewed as the prior-weighted average likelihood. A model's evidence is maximised when its most predictive region of parameter space, i.e. where the prior is highest, coincides with the region of highest likelihood. A model is penalised, however, in two key scenarios: first, a direct conflict where the data favour parameter values the prior deemed unlikely; and second, a penalty for complexity, where a wide prior dilutes the evidence by spreading its predictive power too thinly. Since the prior
must integrates to unity, a broader prior implies a lower height, reducing the evidence integral even if the data are well-fit within that space. The evidence thus naturally rewards models that are both predictive and simple, the latter naturally and quantitatively implements Occam's Razor\footnote{Among competing hypotheses, the one with the fewest assumptions should be selected.}.

A common approach is to make no prior assumption. Models have uniform prior $\prior=\Prob(\model_i)=\mathrm{constant}$, i.e. $\prior_i=\prior_j$, and $\Prob(\model_i|\data)$ simplifies to the ratio of just the evidence of model $\model_i$ to the sum of evidences of all models under comparison,
\begin{equation}
    \Prob(\model_i|\data) = \frac{\evidence_i}{\displaystyle\sum_j \evidence_j}.
\label{eq:model_prob}
\end{equation}
While the Bayes Factor is widely used for comparing two models, the approach in \Cref{eq:model_prob} provides the advantages of yielding the normalised posterior probability for each model, which is not limited to pairwise comparisons and provides an intuitive ranking among the entire set of competing models. We therefore adopt this method for the analysis in \Cref{ssec:model_comparison_results}. Using the evidence values computed from nested sampling in \texttt{unimpeded}, we calculate model probabilities for all eight cosmological models across all datasets, revealing which model extensions are preferred by individual probes and their combinations.

\subsection{Tension Quantification}
\label{ssec:tension_quant_theory}
Tension quantification assesses the statistical consistency between different datasets, say $\data_A$ and $\data_B$, when interpreted under a common underlying model $\model$. In a Bayesian context, several metrics can be employed to diagnose and quantify the degree of agreement or disagreement. The following sections describe five such statistics utilised in this work, drawing on established methods from the literature~\cite{Handley_dimensionality_2019,Lemos2021,Lemos2021TensionMetrics}. For readers interested in alternative approaches to quantifying tensions, the DES collaboration paper~\cite{Lemos2021TensionMetrics} provides a comprehensive comparison of different tension metrics and their applications to cosmological data. The pre-computed chains from \texttt{unimpeded} enable application of any preferred tension metric beyond those presented here. We apply these five complementary statistics to quantify tensions across all pairwise dataset combinations in our grid, with comprehensive results presented in~\Cref{ssec:tension_quantification_results}.
\subsubsection{Combining Likelihoods}
To perform a joint analysis of two statistically independent datasets, $\data_A$ and $\data_B$, their likelihoods are combined multiplicatively: $\likelihood_{AB} = \likelihood_A \likelihood_B$. The posteriors and evidences for the individual and joint datasets are defined as:
\begin{gather}
    \posterior_A = \frac{\likelihood_A\prior_A}{\evidence_A},
    \quad
    \posterior_B = \frac{\likelihood_B\prior_B}{\evidence_B},
    \quad
    \posterior_{AB} = \frac{\likelihood_A\likelihood_B\prior_{AB}}{\evidence_{AB}}.
    \label{eqn:Pdef}
    \\
    \evidence_A = \int\likelihood_A\prior_A\,d{\params},
    \quad
    \evidence_B = \int\likelihood_B\prior_B\,d{\params},\nonumber\\
    \evidence_{AB} = \int\likelihood_A\likelihood_B\prior_{AB}\,d{\params}.
    \label{eqn:Zdef}
\end{gather}
Here, $\prior_A$, $\prior_B$, and $\prior_{AB}$ denote the prior distributions for the individual and joint analyses. In this work, we assume that the priors agree on the shared parameters~\cite{2022arXiv220711457B}, such that $\prior_A(\params_{\text{shared}}) = \prior_B(\params_{\text{shared}}) = \prior_{AB}(\params_{\text{shared}})$, while nuisance parameters unique to each dataset retain their respective priors. $\params$ is taken to be the complete set for the joint analysis, including all cosmological parameters and any nuisance parameters unique to each dataset.

\subsubsection{The \texorpdfstring{$R$}{R} Statistics}
\label{sssec:r_statistic}
The $R$ statistic quantifying the consistency between two datasets, denoted by subscripts $A$ and $B$, within a shared underlying model $\model$~\cite{Marshall2006}. It is defined through a series of equivalent expressions that relate the evidences and conditional probabilities of the datasets:
\begin{equation}
    R = \frac{\evidence_{AB}}{\evidence_A \evidence_B} = \frac{\Prob(\data_A, \data_B)}{\Prob(\data_A)\Prob(\data_B)} = \frac{\Prob(\data_A|\data_B)}{\Prob(\data_A)} = \frac{\Prob(\data_B|\data_A)}{\Prob(\data_B)}.
    \label{eq:R_statistic_full}
\end{equation}
$R$ provides a direct measure of inter-dataset consistency, interpreted with respect to unity. If $R \gg 1$, knowledge of one dataset has strengthened our confidence in the other by a factor of $R$, indicating concordance. If $R \ll 1$, the datasets are inconsistent. The introduction of the second dataset diminishes our confidence in the first under the assumed model, prompting a re-evaluation of the shared model or the datasets themselves. While this establishes a clear framework for consistency, it is crucial to remember that the magnitude of $R$ does not represent an absolute degree of tension, as its value is always conditional on the chosen model and prior.

The $R$ statistic satisfies several desirable properties for a tension metric: it is dimensionally consistent, symmetric with respect to the datasets ($R_{AB}=R_{BA}$), invariant under reparametrisation, and constructed from fundamental Bayesian quantities. However, a crucial property of $R$ is its strong dependence on the prior probability distribution, $\prior(\params)$~\cite{Handley2019}. This dependency can be made explicit by rewriting $R$ in terms of the posteriors $\posterior_A$ and $\posterior_B$:
\begin{equation}
\begin{aligned}
    R &= \frac{1}{\evidence_A \evidence_B} \int \likelihood_A(\params) \likelihood_B(\params) \prior(\params) d\params \\
      &= \int \frac{\likelihood_A(\params) \prior(\params)}{\evidence_A} \frac{\likelihood_B(\params) \prior(\params)}{\evidence_B} \frac{1}{\prior(\params)} d\params \\
      &= \int \frac{\posterior_A(\params) \posterior_B(\params)}{\prior(\params)} d\params \\
      &= \bigg\langle \frac{\posterior_B}{\prior} \bigg\rangle_{\posterior_A} =  \bigg\langle \frac{\posterior_A}{\prior} \bigg\rangle_{\posterior_B},
\end{aligned}
    \label{eq:R_prior_dependence}
\end{equation} 
where we have assumed the datasets are independent. The final line shows that $R$ can be thought of as the posterior average of the ratio of one posterior to the shared prior, averaged over the other posterior.

One should note that reducing the width of the prior on shared, constrained parameters will reduce the value of $R$, thereby increasing the apparent tension between the datasets. This behaviour is opposite to the prior's effect on evidence alone, where narrower priors typically increase the evidence. This creates an attractive balance: one cannot arbitrarily tune priors to increase evidence for a model without simultaneously making it more susceptible to tension if the datasets are not in perfect agreement. While this prior dependence is a feature of a coherent Bayesian analysis, it means that the interpretation of a single $R$ value requires care. If $R$ indicates discordance, this conclusion is robust, since the prior volume effect typically acts to increase $R$ and mask tension. However, if $R$ indicates agreement, one must consider if this is merely the result of an overly wide prior. 

\subsubsection{The Information Ratio}
\label{sssec:information_ratio}
The information ratio, $Q$, is defined in terms of the Kullback-Leibler divergences $\KL$ from the individual data ($\data_A,\data_B$) and joint data ($\data_{AB}$) analyses~\cite{Handley2019}:
\begin{equation}
    Q = \KL^A + \KL^B - \KL^{AB}.
\label{eq:information_ratio}
\end{equation}
To understand the behaviour of $Q$, we can employ the volumetric approximation of~\Cref{eq:kl_divergence_def}, $\KL \approx \log(\vprior) - \log(\vposterior)$, as discussed in~\Cref{sssec:kl_divergence}. Substituting this into~\Cref{eq:information_ratio} yields:
\begin{equation}
    Q \approx \log(\vprior) - \log(\vposterior^A) - \log(\vposterior^B) + \log(\vposterior^{AB}).
\label{eq:logI_volume_approx}
\end{equation}
In a Bayesian interpretation of probability, a highly improbable event is a highly surprising one. $Q$ quantifies this ``surprise'' of agreement, i.e. different datasets making the same predictions. Considering the $\log(\vprior)$ term, a larger prior volume $\vprior$ signifies greater initial uncertainty, making the subsequent agreement of two constraining datasets (with small posterior volumes $\vposterior^A$ and $\vposterior^B$) a more surprising outcome, which results in a larger value of $Q$ and mathematically encodes this greater degree of surprise. In addition, a more constraining dataset results in a smaller $\vposterior$, making $\log(\vposterior)$ a larger negative number and thus $-\log(\vposterior)$ a larger positive number. Consequently, the terms $-\log(\vposterior^A)$ and $-\log(\vposterior^B)$ increase the value of $Q$. This is intuitive: if two highly constraining posteriors (tiny $\vposterior$) end up agreeing, it is far more surprising than if two vague, less constraining posteriors (large $\vposterior$) happen to agree. Conversely, if the datasets are in tension, their posteriors barely overlap, causing the joint posterior volume $\vposterior^{AB}$ to become extremely small. This makes $\log(\vposterior^{AB})$ a large negative number, which in turn significantly decreases the value of $Q$. Therefore, a very low or negative $Q$ is a strong signal of dataset disagreement.

\subsubsection{Suspiciousness}
\label{sssec:suspiciousness}
While the information ratio quantifies the surprise of agreement, the suspiciousness $S$ quantifies the statistical conflict between the likelihoods $\likelihood$ of the two datasets. It is defined by both the prior-dependent information ratio $Q$ and the $R$ statistic~\cite{Handley2019}, and has been applied to quantify tensions in various cosmological contexts~\cite{Joudaki2017MNRAS,Lemos2021}:
\begin{equation}
    \log S = \log R - Q.
\label{eq:suspiciousness_def}
\end{equation}
Since $R$ and $Q$ transform similarly under prior volume alterations, $S$ is largely unaffected by changing the prior widths, as long as this change does not significantly alter the posterior. However, this prior-independence comes at the cost of the direct probabilistic interpretation inherent in $R$, requiring more care to calibrate its scale for significance.
Substituting the definitions of $\log R$ and $Q$ from~\Cref{eq:R_statistic_full,eq:logZ_KL} and~\Cref{eq:information_ratio} into~\Cref{eq:suspiciousness_def}, $\log S$ can be expressed directly in terms of posterior-averaged log-likelihoods:
\begin{equation}
    \log S = \langle \log{\likelihood_{AB}} \rangle_{\posterior_{AB}} - \langle \log{\likelihood_{A}} \rangle_{\posterior_{A}} - \langle \log{\likelihood_{B}} \rangle_{\posterior_{B}}.
\label{eq:suspiciousness_likelihood_avg}
\end{equation}
When likelihoods $\likelihood_A$ and $\likelihood_B$ are in strong agreement, $\log S$ is zero or positive. Conversely, if the likelihoods are in tension, $\log S$ becomes negative, with larger negative values indicating stronger tension. $\log S$ can also be calibrated into a tension probability, $p$, and an equivalent significance in Gaussian standard deviations, $\sigma$ (see~\Cref{sssec:p_and_sigma}).

\subsubsection{Bayesian Model Dimensionality}
\label{sssec:bayesian_model_dimensionality}
While the Kullback-Leibler divergence, $\KL$, discussed in~\Cref{sssec:kl_divergence} provides a single value for the total information gain, it marginalises out any information about individual parameters. It cannot tell us how many parameters are being constrained by the data, nor what each parameter is constraining. For instance, a strong, correlated constraint between two parameters can yield the same $\KL$ as two well-constrained but independent parameters (visually demonstrated in~\cite{Handley_dimensionality_2019}). In high-dimensional cosmological analyses, where corner plots~\cite{2016JOSS....1...24F} show only marginalised views and can hide complex degeneracies, a metric is needed to quantify the effective number of constrained parameters. The Bayesian Model Dimensionality, $d$, was introduced to fulfil this role~\cite{Handley_dimensionality_2019} and is defined as:

\begin{equation}
    \begin{split}
    \frac{d}{2} &= \int \posterior(\params) \left(\log\frac{\posterior(\params)}{\prior(\params)} - \KL\right)^2 d\params \\
    & = \left\langle{\left(\log\frac{\posterior}{\prior}\right)}^2\right\rangle_{\posterior} - {\left\langle\log\frac{\posterior}{\prior}\right\rangle}_{\posterior}^2 \\
    & = \mathrm{var}(\shannon)_{\posterior}  \\
    & = \langle(\log \likelihood)^2\rangle_{\posterior} - \langle\log\likelihood\rangle_{\posterior}^2,
    \end{split}
\label{eq:bayesian_dimensionality}
\end{equation}
where $\shannon = \log\frac{\posterior}{\prior}$ is the Shannon Information mentioned in~\Cref{eq:shannon_info}. $d$ is the variance of $\shannon$ over posterior, and hence, a higher-order statistic than the KL divergence.The Bayesian Model Dimensionality possesses several important properties. Crucially, it is only weakly prior-dependent, as the evidence contributions required to normalise the posterior and prior in the $\shannon$ term in~\Cref{eq:bayesian_dimensionality} effectively cancel out. Furthermore, like the KL divergence, it is additive for independent parameters and invariant under a change of variables. When combining datasets,the number of parameters that are constrained becomes:
\begin{equation}
    d_{A \cap B} = d_A + d_B - d_{AB}.
\end{equation}

\subsubsection{Tension Probability and Significance}
\label{sssec:p_and_sigma}
Suspiciousness discussed in~\Cref{sssec:suspiciousness} can be calibrated into a more intuitive tension probability, $p$, and an equivalent significance expressed in Gaussian standard deviations, $\sigma$. This calibration relies on the approximation that in the case of a Gaussian likelihood, the quantity $d - 2\log S$ follows a $\chi^2$ distribution. The number of degrees of freedom is given by the Bayesian model dimensionality, $d$, as discussed in~\Cref{sssec:bayesian_model_dimensionality}. $p$ represents the probability that a level of discordance at least as large as the one observed could arise by chance. It is calculated using the survival function of the $\chi_d^2$ distribution:
\begin{equation}
  p = \int_{d-2\log S}^{\infty} \chi_d^2(x)\,\mathrm{d}x = \int_{d-2\log S}^{\infty} \frac{x^{d/2-1}e^{-x/2}}{2^{d/2}\Gamma(d/2)}\,\mathrm{d}x.
\label{eq:tension_probability}
\end{equation}
This $p$-value can then be converted into an equivalent significance on a Gaussian scale, $\sigma$, using the inverse complementary error function ($\mathrm{Erfc}^{-1}$):
\begin{equation}
    \sigma = \sqrt{2}\,\mathrm{Erfc}^{-1}(p).
\label{eq:sigma_conversion}
\end{equation}
Following standard conventions, if $p \lesssim 0.05$ (corresponding to $\sigma \gtrsim 2$), the datasets are considered to be in moderate tension, while $p \lesssim 0.003$ ($\sigma \gtrsim 3$) corresponds to strong tension.

\subsubsection{The Look Elsewhere Effect}
\label{sssec:look_elsewhere_effect}
The Look Elsewhere Effect (LEE) arises when multiple statistical tests are performed, increasing the probability of finding a seemingly significant result purely by chance. This effect is particularly relevant to our analysis, where we systematically evaluate tension statistics for $N=248$ distinct model-dataset combinations~(see \Cref{sec:results}). Without accounting for multiple comparisons, the likelihood of encountering at least one false positive becomes substantial.

Rather than applying a Bonferroni correction to each individual $p$-value (which would change as the grid expands), we instead adopt a significance threshold that naturally accounts for the look elsewhere effect. Under the null hypothesis of no genuine tension, $p$-values are uniformly distributed between 0 and 1. Therefore, if we perform $N=248$ independent tests, we expect exactly one result to have $p \leq 1/N$ purely by chance. This provides a natural threshold:
\begin{equation}
    \sigma_{\text{threshold}} = \sqrt{2}\,\mathrm{Erfc}^{-1}\left(\frac{1}{N}\right).
\label{eq:sigma_threshold_corrected}
\end{equation}
For our grid with $N=248$, this gives $\sigma_{\text{threshold}} \approx 2.88$. This threshold is not arbitrary, it represents the significance level at which we would expect only one false positive across all 248 tests if there were no genuine tensions. Any result highlighted above this threshold is more extreme than what we would expect from random fluctuations alone. This approach has the advantage that the threshold itself reflects the scope of the analysis, while individual $p$-values and $\sigma$ values remain interpretable independently of the grid size.

\subsubsection{Model-Weighted Average Tension Statistics}
\label{sssec:model_weighted_average}
To evaluate the overall tension between datasets across our entire model space, we employ a model-weighted average for each tension statistic. This approach provides a single, summary ranking of datasets that accounts for the fact that some models are better supported by the data than others, and therefore, their tension statistics should have a heavier weighting under the Bayesian framework. The tension statistic for each model is weighted by its posterior probability, $\Prob(\model_i|\data)$, calculated by~\Cref{eq:model_posterior}. For a generic tension metric between datasets $\data_A$ and $\data_B$, this average is computed as:
\begin{equation}
    \langle \text{Statistic}(\data_A, \data_B) \rangle_\model = \displaystyle\sum_{i} \Prob(\model_i|\data) \times \text{Statistic}(\model_i, \data_A, \data_B).
    \label{eq:model_weighted_average}
\end{equation}
For example, the model-weighted Kullback-Leibler divergence is:
\begin{equation}
    \langle \KL \rangle_{\Prob(\model)} = \displaystyle\sum_{i} \Prob(\model_i|\data) \times \KL(\model_i).
    \label{eq:model_weighted_kl}
\end{equation}
Similarly, the model-weighted tension significance is:
\begin{equation}
    \langle \sigma \rangle_{\Prob(\model)} = \displaystyle\sum_{i} \Prob(\model_i|\data) \times \sigma(\model_i).
    \label{eq:model_weighted_sigma}
\end{equation}
These model-weighted statistics are presented in the heatmaps in~\Cref{ssec:model_comparison_results,ssec:tension_quantification_results}.

\subsubsection{Prior and Model Dependence of Tension Metrics}
\label{sssec:prior_model_dependence}
It is important to stress that while these metrics provide a valuable quantitative framework for tension quantification, they are inherently conditional on the assumptions of the analysis. Specifically, the Bayesian evidence, and by extension the $R$-value and information ratio, are sensitive to the chosen prior volume for the model parameters. Similarly, all metrics are fundamentally model-dependent, as they compare datasets or models within a predefined theoretical space. Consequently, the numerical values reported in this paper should not be interpreted as absolute, model-independent measures of tension or evidence. Instead, they represent the relative preference of the data for different hypotheses under a consistent and explicitly stated set of prior and model assumptions. For recent discussions of prior dependence in the context of CMB lensing tensions, see e.g.~\cite{2023arXiv231008490O}. The treatment of cosmological and nuisance parameters that underlies these metrics is set out in detail in~\Cref{sssec:treatment_unconstrained_nuisance}.

\subsection{Cosmological Models}
\label{ssec:cosmological_models}
We consider a comprehensive set of cosmological models extending the standard $\Lambda$CDM paradigm. All models are described within the framework of the Friedmann-Lemaître-Robertson-Walker (FLRW) metric with general relativity. The background expansion is governed by the Friedmann equation, and initial conditions are set by nearly scale-invariant, adiabatic Gaussian scalar perturbations~\cite{Planck2013params}. All models are implemented using the Cobaya framework~\cite{cobayaascl,Torrado2021Cobaya}, which interfaces with the CAMB Boltzmann code~\cite{Lewis:1999bs} to compute theoretical predictions for the observables. Each model is elaborated in detail in the following subsubsections.

\subsubsection{Baseline: $\Lambda$CDM}
\label{sssec:lcdm}
The baseline $\Lambda$CDM model describes a spatially flat universe with a cosmological constant (dark energy equation of state $w = -1$), cold dark matter, and a power-law spectrum of adiabatic scalar perturbations. The cold dark matter paradigm was established by~\cite{Blumenthal1984}, while the cosmological constant component emerged from the discovery of cosmic acceleration through Type Ia supernovae observations~\cite{Riess1998,Perlmutter1999}. The model has been extensively validated by cosmic microwave background measurements~\cite{Planck2013params}. The model is characterized by six fundamental parameters: the baryon density parameter $\omega_b = \Omega_b h^2$, the cold dark matter density parameter $\omega_c = \Omega_c h^2$, the angular scale of the sound horizon at recombination $\theta_*$ (often parameterised as $100\theta_{MC}$), the reionisation optical depth $\tau$, the scalar spectral index $n_s$, and the amplitude of scalar perturbations $\ln(10^{10}A_s)$.

The Hubble parameter evolves as:
\begin{equation}
    H^2(a) = H_0^2 \left[ \Omega_r a^{-4} + \Omega_m a^{-3} + \Omega_\Lambda \right],
    \label{eq:friedmann_lcdm}
\end{equation}
where $a$ is the scale factor, $\Omega_r$ is the radiation density parameter, $\Omega_m$ is the total matter density parameter, and $\Omega_\Lambda$ is the cosmological constant density parameter. The flatness constraint imposes $\Omega_r + \Omega_m + \Omega_\Lambda = 1$.

The primordial scalar power spectrum is:
\begin{equation}
    \mathcal{P}_s(k) = A_s \left( \frac{k}{k_0} \right)^{n_s - 1},
    \label{eq:power_spectrum_lcdm}
\end{equation}
where $k_0 = 0.05\,\text{Mpc}^{-1}$ is the pivot scale. The baseline assumes three standard neutrinos with $N_{\text{eff}} = 3.046$ and minimal neutrino mass $\Sigma m_\nu = 0.06\,\text{eV}$. The helium abundance is computed consistently with Big Bang nucleosynthesis.

\textbf{Free parameters:} $\omega_b$, $\omega_c$, $\theta_*$, $\tau$, $n_s$, $\ln(10^{10}A_s)$.

\subsubsection{Varying Curvature: $\Omega_k\Lambda$CDM}
\label{sssec:klcdm}
This extension allows for spatial curvature by freeing the curvature density parameter $\Omega_k$~\cite{Planck2013params}. The Friedmann equation becomes:
\begin{equation}
    H^2(a) = H_0^2 \left[ \Omega_r a^{-4} + \Omega_m a^{-3} + \Omega_k a^{-2} + \Omega_\Lambda \right],
    \label{eq:friedmann_klcdm}
\end{equation}
with the constraint $\Omega_r + \Omega_m + \Omega_k + \Omega_\Lambda = 1$. Positive $\Omega_k$ corresponds to an open universe (negative spatial curvature), while negative $\Omega_k$ describes a closed universe (positive spatial curvature). Non-zero curvature alters the universe's spatial geometry, modifying the evolution of metric perturbations and photon geodesics. Boltzmann codes account for this by solving the perturbation equations on a curved background, which causes a characteristic angular shift in the CMB power spectrum's acoustic peaks and modifies the late-time integrated Sachs-Wolfe effect.

\textbf{Free parameters:} $\Lambda$CDM parameters + $\Omega_k$.

\subsubsection{Constant Dark Energy Equation of State: $w$CDM}
\label{sssec:wcdm}
This model generalizes the cosmological constant to a dark energy fluid with constant equation of state $w = p_{\text{DE}}/\rho_{\text{DE}}$~\cite{Turner1997}. The dark energy density evolves as $\rho_{\text{DE}}(a) \propto a^{-3(1+w)}$, modifying the Friedmann equation to:
\begin{equation}
    H^2(a) = H_0^2 \left[ \Omega_r a^{-4} + \Omega_m a^{-3} + \Omega_{\text{DE}} a^{-3(1+w)} \right].
    \label{eq:friedmann_wcdm}
\end{equation}
The model assumes spatial flatness. Within the Parameterised Post-Friedmann (PPF) framework, dark energy is treated as a perfect fluid with a sound speed typically set to unity ($c_s^2=1$). Consequently, dark energy perturbations are negligible on sub-horizon scales, influencing structure growth primarily through the modified background expansion and the late-time integrated Sachs-Wolfe (ISW) effect on the CMB.

\textbf{Free parameters:} $\Lambda$CDM parameters + $w$.

\subsubsection{Time-Varying Dark Energy: $w_0w_a$CDM}
\label{sssec:w0wacdm}
This model allows for time-varying dark energy using the Chevallier-Polarski-Linder (CPL) parameterisation~\cite{Chevallier2001,Linder2003}:
\begin{equation}
    w(a) = w_0 + w_a(1-a),
    \label{eq:cpl_parameterisation}
\end{equation}
where $w_0$ is the present-day equation of state and $w_a$ characterizes its time evolution. The dark energy density evolves as:
\begin{equation}
    \rho_{\text{DE}}(a) = \rho_{\text{DE},0} \, a^{-3(1+w_0+w_a)} \exp[-3w_a(1-a)].
    \label{eq:rho_de_cpl}
\end{equation}
The model assumes spatial flatness. Similar to $w$CDM, this model uses the PPF formalism where dark energy is a smooth component with $c_s^2=1$, preventing it from clustering. The time-varying equation of state produces a more complex background evolution, altering the growth history of matter perturbations and creating a distinct signature in the late-time ISW effect compared to a constant $w$.

\textbf{Free parameters:} $\Lambda$CDM parameters + $w_0$ + $w_a$.

\subsubsection{Varying Lensing Amplitude: $A_L\Lambda$CDM}
\label{sssec:alcdm}
This model introduces a phenomenological parameter $A_L$ that scales the lensing potential power spectrum, allowing for deviations from the standard lensing predictions~\cite{Planck2013params}. The parameter modifies the lensed CMB power spectra by scaling the lensing potential correlations:
\begin{equation}
    C_\ell^{\text{lensed}} = C_\ell^{\text{unlensed}} + A_L \Delta C_\ell^{\text{lensing}},
    \label{eq:lensing_amplitude}
\end{equation}
where $\Delta C_\ell^{\text{lensing}}$ represents the correction due to gravitational lensing. The standard $\Lambda$CDM prediction corresponds to $A_L = 1$. Values $A_L > 1$ indicate enhanced lensing effects, while $A_L < 1$ suggest reduced lensing. This extension was motivated by the Planck collaboration's observation of a preference for $A_L > 1$ in the CMB temperature data, providing a way to test the consistency of gravitational lensing predictions. This parameter does not alter the physical evolution of perturbations but acts as a phenomenological scaling of the gravitational lensing potential. In Boltzmann codes, the calculated lensing potential power spectrum is multiplied by $A_L$, which directly modifies the smoothing of the CMB acoustic peaks and the amplitude of the lensing-induced B-mode spectrum.

\textbf{Free parameters:} $\Lambda$CDM parameters + $A_L$.

\subsubsection{Varying Neutrino Masses: $m_\nu\Lambda$CDM}
\label{sssec:mnulcdm}
This model extends the standard $\Lambda$CDM framework by allowing the sum of the three active neutrino masses, $\Sigma m_\nu$, to vary as a free parameter. The effective number of relativistic species is held fixed at the standard value, $N_{\text{eff}} = 3.046$~\cite{Mangano2005}. The contribution of massive neutrinos to the cosmic energy budget today is:
\begin{equation}
    \Omega_\nu h^2 = \frac{\Sigma m_\nu}{93.14\,\text{eV}}.
    \label{eq:neutrino_density}
\end{equation}
Massive neutrinos act as hot dark matter, suppressing the growth of structure below their free-streaming length due to their large thermal velocities. Boltzmann codes solve the full neutrino Boltzmann equation to model this effect, which manifests as a distinct, scale-dependent suppression in the matter power spectrum and subtly alters the CMB via the early ISW effect.

\textbf{Free parameters:} $\Lambda$CDM parameters + $\Sigma m_\nu$.

\subsubsection{Running Spectral Index: $n_{\text{run}}\Lambda$CDM}
\label{sssec:running_lcdm}
This model extends the primordial power spectrum beyond a simple power law by allowing the scalar spectral index $n_s$ to vary with scale $k$. This scale dependence is parameterised by the ``running of the spectral index,'' defined as $n_{\text{run}} \equiv dn_s/d\ln k$~\cite{Kosowsky1995,Planck2013params}. While the baseline $\Lambda$CDM model assumes $n_{\text{run}} = 0$, a non-zero running is a generic prediction of many inflationary models. In the context of single-field slow-roll inflation, the running is a second-order effect in the slow-roll parameters and is predicted to be very small. A detection of a significant non-zero running would therefore challenge the simplest inflationary scenarios and provide crucial insights into the shape of the inflaton potential or point towards more complex physics in the early universe.

The primordial scalar power spectrum is modified to include a logarithmic scale-dependent term in the exponent:
\begin{equation}
    \mathcal{P}_s(k) = A_s \left( \frac{k}{k_0} \right)^{n_s - 1 + \frac{1}{2}n_{\text{run}} \ln\left(\frac{k}{k_0}\right)},
    \label{eq:running_spectrum}
\end{equation}
where $n_s$ is the spectral index and $n_{\text{run}}$ is the running, both evaluated at the pivot scale $k_0 = 0.05\,\text{Mpc}^{-1}$. This form arises from a first-order Taylor expansion of the spectral index $n_s(k)$ in $\ln k$.

\textbf{Free parameters:} $\Lambda$CDM parameters + $n_{\text{run}}$.

\subsubsection{Primordial Gravitational Waves: $r\Lambda$CDM}
\label{sssec:r_lcdm}
This model includes primordial tensor perturbations (gravitational waves) characterized by the tensor-to-scalar ratio $r$ at the pivot scale, providing a key test of inflationary theory~\cite{Guth1981,Starobinsky1980}. The spectrum of relic gravitational waves from inflation was first calculated by~\cite{Starobinsky1979}. The tensor power spectrum is:
\begin{equation}
    \mathcal{P}_t(k) = A_t \left( \frac{k}{k_p} \right)^{n_t},
    \label{eq:tensor_spectrum}
\end{equation}
where $r = A_t/A_s$ is evaluated at a chosen pivot scale (typically $k_p = 0.002\,\text{Mpc}^{-1}$ for tensor modes). The tensor spectral index $n_t$ is often constrained by the inflationary consistency relation $n_t = -r/8$.

\textbf{Free parameters:} $\Lambda$CDM parameters + $r$ (and optionally $n_t$ if not fixed by consistency relation).

\section{Methodology}
\label{sec:methodology}
\subsection{Nested Sampling}
\label{ssec:nested_sampling}
Nested sampling, introduced by Skilling (2004)~\cite{Skilling2004,Skilling2006}, is a Monte Carlo method designed for Bayesian computation. It is particularly powerful for calculating the Bayesian evidence (or marginal likelihood), a key quantity for model comparison and tension quantification, while simultaneously producing posterior samples for parameter estimation. The algorithm transforms the multi-dimensional evidence integral into a one-dimensional integral over prior volume, which is then solved numerically.

Nested sampling is generally considered the ``ground truth'' method for evidence calculation, representing the reference standard against which other approaches are compared. While alternative methods exist that aim to calculate evidence more efficiently, such as harmonic mean estimators~\cite{Piras2024harmonic} and MC evidence~\cite{Heavens2017MCEvidence}, these typically require validation against nested sampling results to establish their accuracy. The pre-computed grid of nested sampling chains provided by \texttt{unimpeded} therefore serves not only as a resource for cosmological model comparison and tension quantification, but also as a reference dataset for assessing the performance of alternative evidence estimation techniques. Recent comprehensive reviews of nested sampling methodology and applications can be found in~\cite{Buchner2023,Ashton2022NRvMP}.

We use the publicly available \texttt{PolyChord} sampler~\cite{Handley2015PolychordI,Handley2015PolychordII}, which provides a robust and efficient implementation of nested sampling well suited to the high-dimensional parameter spaces typical of modern cosmology. This section outlines the core methodology of the nested sampling algorithm.

\subsubsection{Generating Samples and Increasing Likelihood}
\label{sssec:nested_sampling_algorithm}
The fundamental principle of nested sampling is to explore the parameter space $\params$ by iteratively moving through nested contours of constant likelihood $\likelihood_i$. The algorithm begins by drawing $n_0$ initial ``live points'' from the prior distribution $\pi(\params)$. At each iteration $i$, the point with the lowest likelihood, $\likelihood_i$, among the current set of live points is identified. By some criterion we then choose whether or not to remove this from the live set and add it to a collection of ``dead points''. By a different criterion it is then replaced with new points drawn from the prior $\pi(\params)$, subject to the hard constraint that their likelihood $\likelihood(\params)$ must be greater than $\likelihood_i$. This ensures that the likelihoods of the dead points, $\{\likelihood_1, \likelihood_2, \likelihood_3, \dots\}$, form a monotonically increasing sequence.

\subsubsection{Prior Volume Contraction}
This iterative deletion of live points with the lowest likelihood systematically contracts the region of parameter space, leading to the peak(s) of the posterior. The prior mass $X(\likelihood)$ is the measure of the fraction of the prior mass\footnote{In the context of nested sampling, ``prior volume'' and ``prior mass'' are used interchangeably to refer to the same fundamental concept.} contained within an iso-likelihood contour $\likelihood(\params) = \likelihood$. It is calculated by integrating the element of prior mass $dX = \prior(\params)d\params$ covering all likelihood values greater than $\likelihood$~\cite{Skilling2006}:
\begin{equation}
    X(\likelihood) = \int_{\likelihood(\params)>\likelihood} \pi(\params) d\params.
    \label{eq:prior_volume}
\end{equation}
By construction described in~\Cref{sssec:nested_sampling_algorithm}, the algorithm generates a sequence of increasing likelihoods $\likelihood_1 < \likelihood_2 < \dots < \likelihood_i$ corresponding to a sequence of shrinking prior volumes $X_i = X(\likelihood_i)$, where $X_1 > X_2 > \dots > X_i$. At each iteration $i$, the removal of the point with likelihood $\likelihood_i$ corresponds to shrinking the prior volume from $X_{i-1}$ to $X_i$, so the ratio of the volumes $t_i = X_i/X_{i-1}$. The initial prior volume is $X_0 = 1$ (the entire prior).

More formally, the shrinkage of the prior volume is a stochastic process with distribution ${P(t_i) = n_i t_i^{n_i-1}}$, where $n_i$ is the live point count at iteration $i$. The expected logarithm of the prior volume at iteration $i$ is given by the sum over the live point count $n_k$ at each preceding iteration $k$~\cite{Hu2023aeons}:
\begin{equation}
    \langle \log X_i \rangle = -\sum_{k=1}^{i} \frac{1}{n_k}.
    \label{eq:logX_general}
\end{equation}
In the simplified case of a constant number of live points, $n_k = n_{\text{live}}$ for all $k$, this sum reduces to $\langle \log X_i \rangle = -i/n_{\text{live}}$. This leads to the well-known exponential approximation for the prior volume, $\langle X_i \rangle \approx e^{-i/n_{\text{live}}}$.

This exponential compression allows the algorithm to efficiently traverse the parameter space from the broad prior towards the narrow, high-likelihood regions where the posterior mass is concentrated. However, as we discuss in~\Cref{ssec:dynamic_ns}, \Cref{eq:logX_general} provides the rigorous framework necessary for analysing runs where the number of live points varies. Each dead point is associated with a specific set of parameters $\params_i$, likelihood $\likelihood_i$ and an estimated prior volume $X_i$, enabling the reconstruction of the evidence integral, discussed in~\Cref{sssec:evidence_estimation}.

\subsubsection{Evidence Estimation}
\label{sssec:evidence_estimation}
The primary strength of nested sampling is its ability to directly calculate the Bayesian evidence, $\evidence = \int \likelihood(\params) \pi(\params) d\params$. Instead of integrating the likelihood over all possible parameters, which is computationally prohibitive in high dimensions, this integral can be reformulated in terms of the prior volume $X$. Since $\likelihood$ can be expressed as an inverse function of its enclosed prior volume, $\likelihood(X)$, the evidence integral can be rewritten as a one-dimensional integral from $X=0$ to $X=1$~\cite{Skilling2006}:
\begin{equation}
    \evidence = \int_0^1 \likelihood(X) dX.
\label{eq:evidence_integral}
\end{equation}
The nested sampling algorithm provides a discrete sequence of points $(\likelihood_i, X_i)$ that allows for a numerical approximation of this integral. Using a simple quadrature scheme, the evidence can be estimated as a weighted sum over the discarded ``dead'' points:
\begin{equation}
    \evidence \approx \displaystyle\sum_{i \in \text{dead}} w_i \likelihood_i,
\label{eq:evidence_sum}
\end{equation}
where $\likelihood_i$ is the likelihood of the $i$-th discarded point and $w_i$ is the associated prior volume, or weight. This weight represents the prior mass contained within the shell between successive likelihood contours, $w_i = X_{i-1} - X_i$. In practice, we approximate $X_i \approx e^{-i/n_{\text{live}}}$ and can thereby estimate the weights for the summation. The evidence can then be used in model comparison and tension quantification.

\subsubsection{Importance Weight and Posterior Estimation}
In addition to evidence calculation, the collection of live and dead points can be used to derive posterior inferences, and hence, for parameter estimation. Each ``dead point'' $\params_i$ is associated with a likelihood $\likelihood_i$ and a prior mass weight, $w_i$, which represents the element of prior mass of the shell in which point $\params_i$ was sampled. The importance weight, or the posterior, $p_i$ for each dead point is its contribution to the evidence $\evidence_i = w_i \likelihood_i$, normalised by the total evidence $\evidence$ from~\Cref{eq:evidence_sum}:
\begin{equation}
    p_i = \frac{w_i \likelihood_i}{\evidence}.
\label{eq:importance_weight}
\end{equation}

The posterior expectation value for a function of the parameters, $f(\params)$, is the weighted sum:
\begin{equation}
    \langle f(\params) \rangle \approx \displaystyle\sum_{j \in \{\text{dead}\}} p_j f(\params_j).
\end{equation}
This allows for the construction of marginalised posterior distributions, credibility intervals, and other standard Bayesian parameter summaries.

\subsubsection{Algorithm Termination and Stopping Criterion}
As iterations repeat, prior mass weights $w_i$ monotonically decrease, and the likelihoods $\likelihood_i$ monotonically increase. Live points are therefore concentrated in regions of high likelihood, and are associated with tiny prior mass. The nested sampling algorithm is terminated at the $i$-th iteration when the remaining posterior mass is some small fraction of the currently calculated evidence:
\begin{equation}
    Z_{\text{live}} \approx \langle \likelihood_{\text{live}} \rangle X_i,
    \label{eq:live_evidence}
\end{equation}
where $\langle \likelihood_{\text{live}} \rangle$ is the average likelihood of the current live points. By this stage, the estimated remaining evidence from the live points is a negligible fraction of the evidence accumulated thus far. 
A common stopping criterion is to halt the process when the expected future contribution to the evidence is smaller than a user-defined tolerance $\epsilon$:
\begin{equation}
    Z_{\text{live}} < \epsilon Z_{\text{dead}}.
\end{equation}
The stopping criterion ensures that the  final evidence estimate and posterior samples are robust and that computational effort is not wasted on regions of the parameter space with insignificant posterior mass.

\subsubsection{Evidence Correction for Unphysical Parameter Space}
\label{ssec:evidence_correction}

The standard nested sampling algorithm, as outlined in the preceding sections, implicitly assumes that the entire prior volume is accessible and yields a non-zero likelihood. In practice, many cosmological models possess parameter spaces with regions that are ``unphysical.'' These are regions where the model violates fundamental physical constraints, such as predicting a negative age for the Universe, failing to converge during numerical evolution, or producing spectra with unphysical features. In our analysis pipeline, these unphysical points are assigned a minimal log-likelihood value, effectively a numerical log-zero. We therefore partition the parameter space $\params \in \Omega$ into two disjoint subspaces: the ``physical'' subspace $\Omega_{\text{phys}}$, where the likelihood $\likelihood(\data|\params,\model) > 0$, and the ``unphysical'' subspace $\Omega_{\text{unphys}}$, where $\likelihood(\data|\params,\model) \le 0$. The process of generating samples from the prior $\prior(\params|\model)$ typically begins by drawing a point from a unit hypercube, a $D$-dimensional space $[0,1]^D$, where $D$ is the number of model parameters $\paramsM$, with each coordinate sampled uniformly as $(u_1, u_2, \dots, u_D)$ where $u_i \in [0,1]$, which is then transformed into the physical parameter space via the prior transformation. The physical parameter space corresponds to the actual parameter ranges (e.g., $H_0 \in [60, 80]$ km s$^{-1}$ Mpc$^{-1}$, $\Omega_{\text{b}}h^2 \in [0.019, 0.026]$), obtained by applying the appropriate prior type (uniform, Gaussian, log-uniform, etc.) to each unit hypercube coordinate. This transformed point may fall into either $\Omega_{\text{phys}}$ or $\Omega_{\text{unphys}}$. Since only points in $\Omega_{\text{phys}}$ will enter the nested sampling algorithm and contribute to the evidence integral, it is crucial to account for the fraction of the prior volume that is inaccessible due to unphysicality. This accessible volume fraction can be estimated via rejection sampling:
\begin{equation}
    V_{\text{phys}} \approx \frac{n_{\text{prior}}}{n_{\text{total}}} = \frac{\text{\# of points in } \Omega_{\text{phys}}}{\text{\# of points in } \Omega_{\text{phys}} + \text{\# of points in } \Omega_{\text{unphys}}}.
\end{equation}
The corrected, true evidence is therefore:
\begin{equation}
    \evidence_{\text{true}} = \evidence_{\text{raw}} \times \left( \frac{n_{\text{prior}}}{n_{\text{total}}} \right).
\end{equation}
In logarithmic form, which is used for all computations, the correction is additive:
\begin{equation}
    \log(\evidence_{\text{true}}) = \log(\evidence_{\text{raw}}) + \log\left(\frac{n_{\text{prior}}}{n_{\text{total}}}\right).
    \label{eq:evidence_correction}
\end{equation}
It is important to note that the correction factor is $n_{\text{prior}}/n_{\text{total}}$, not $n_{\text{live}}/n_{\text{total}}$. This is because the volume shrinkage during the initial compression phase (where $n_{\text{prior}}$ is reduced to $n_{\text{live}}$) is already correctly tracked by \texttt{PolyChord}'s \texttt{update\_evidence()} routine at each nested sampling step. The ratio $n_{\text{prior}}/n_{\text{total}}$ accounts only for the fraction of parameter space that is physical versus unphysical, as determined by the initial rejection sampling.

\subsubsection{Dynamic Nested Sampling \& Synchronous Parallel Sampling}
\label{ssec:dynamic_ns}

\begin{figure*}
\centering
\includegraphics[width=\textwidth]{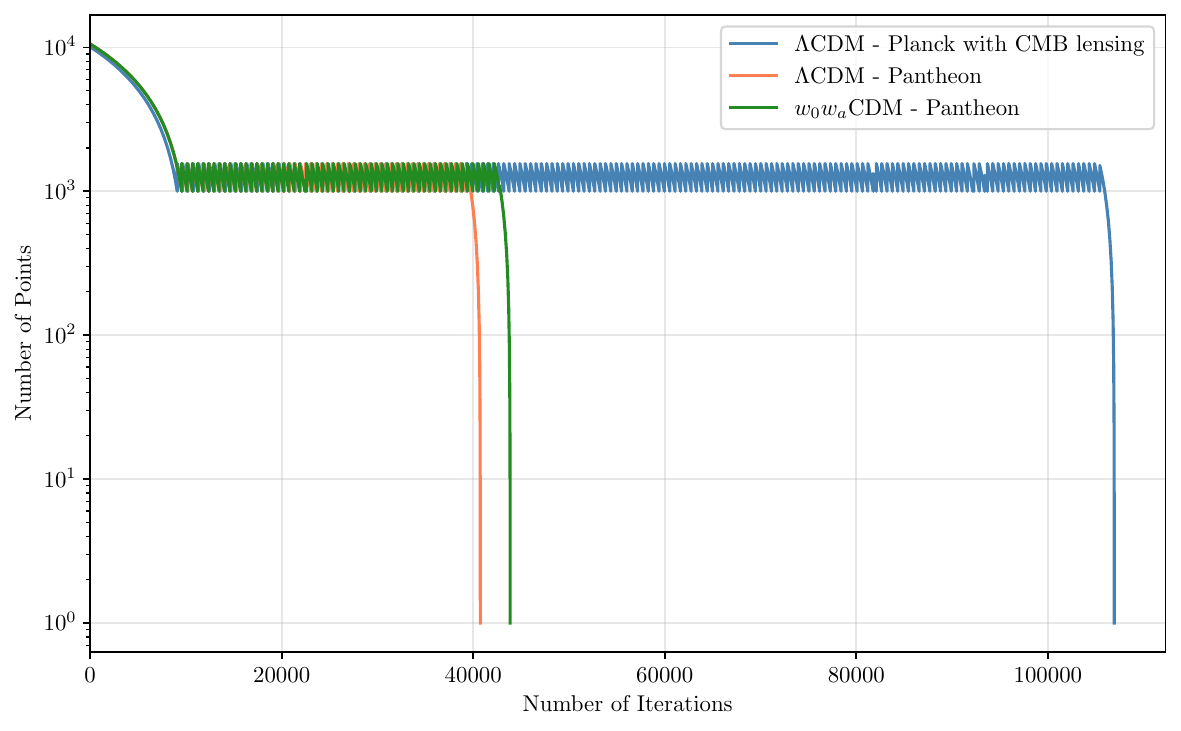}
\caption{Evolution of the live point count throughout nested sampling runs for $\Lambda$CDM and $w_0w_a$CDM models with Planck with CMB lensing and Pantheon datasets. Three distinct phases are visible: (1) initial compression where the first $n_{\text{prior}} - n_{\text{live}} \approx 9{,}000$ iterations sequentially delete the lowest-likelihood points without replacement, reducing the live point count from $n_{\text{prior}} \approx 10^4$ to the target value $n_{\text{live}} \approx 10^3$, which will enter the main nested sampling stage; (2) main sampling phase where the live point count oscillates above $n_{\text{live}}$ due to synchronous parallel processing with 760 cores; (3) final deletion phase where the live point count decreases from $n_{\text{live}}$ to zero as the remaining points are systematically removed one by one. The iteration number $i$ ($x$-axis) maps to the compressed log-prior volume via $\langle \log X_i \rangle = -\sum_{k=1}^{i} 1/n_k$, where $n_k$ is the live point count at iteration $k$, as shown by the $y$-axis (see~\Cref{ssec:dynamic_ns} for details)~\cite{Hu2023aeons}. The different termination points reflect the different Kullback--Leibler divergences between prior and posterior for each model-dataset combination. More complex models like $w_0w_a$CDM (green) require slightly more iterations than simpler models like $\Lambda$CDM (orange), but this effect is not as dominant as the variation across different datasets, with Planck with CMB lensing (blue) requiring substantially more iterations than Pantheon (orange, green). }
\label{fig:dns_plots}
\end{figure*}

Our statistical analysis is performed using the dynamic nested sampling framework implemented in \texttt{PolyChord}~\cite{Handley2015PolychordI,Handley2015PolychordII}. While this framework supports adaptive live point allocation~\cite{Higson2019dns}, we use a constant target number of live points during the main sampling phase, leveraging the framework's efficient synchronous parallelisation across HPC cores and integrated termination scheme. Our runs proceed through three distinct phases, as illustrated in Figure~\ref{fig:dns_plots}.

\textbf{Initial Compression.} The process begins with an initial set of $n_{\text{prior}} \approx 10{,}000$ live points sampled directly from the prior and verified to yield physical solutions (i.e., $\likelihood(\mathcal{D}|\boldsymbol{\theta},\mathcal{M}) > 0$). The compression phase consists of the first $n_{\text{prior}} - n_{\text{live}} \approx 9{,}000$ iterations of the nested sampling algorithm, during which the lowest-likelihood points are sequentially deleted without replacement, reducing the live point count from $n_{\text{prior}}$ to the target value $n_{\text{live}} \approx 1{,}000$, which will enter the main nested sampling stage. During this phase, the number of active points $n_k$ in \Cref{eq:logX_general} decreases from $n_{\text{prior}}$ down to $n_{\text{live}}$. Each deleted point contributes to the evidence integral with its appropriate prior volume weight $w_i = X_{i-1} - X_i$, where the prior volume shrinks according to the decreasing live point count. This phase efficiently accumulates evidence from the vast, low-likelihood regions of the prior volume.

\textbf{Synchronous Parallel Sampling.} During the main sampling phase, the live point count oscillates above the target value $n_{\text{live}} \approx 1{,}000$, corresponding to $n_k \approx n_{\text{live}}$ in \Cref{eq:logX_general}. These oscillations, clearly visible in Figure~\ref{fig:dns_plots}, are a characteristic feature of \texttt{PolyChord}'s synchronous parallelisation scheme. In each iteration, a batch of the lowest-likelihood points (equal to the number of parallel cores, in our case 760) is discarded, and the same number of new points are generated simultaneously. This synchronous approach, where all cores must wait for the slowest likelihood evaluation in the batch to complete, is crucial for preventing statistical bias. An asynchronous approach would preferentially sample regions of parameter space with faster likelihood evaluations (e.g., flat universes over curved ones), leading to an incorrect posterior.

\textbf{Final Deletion.} The run terminates with a final phase where all remaining live points are systematically removed one by one. In this stage, the active point count $n_k$ decreases from $n_{\text{live}}$ down to 1. This process is the dynamic framework's integrated termination procedure, which replaces the separate calculation of the remaining evidence $Z_{\text{live}} \approx \langle \likelihood \rangle_{\text{live}} X_i$ found in earlier nested sampling implementations~\cite{Handley2015PolychordII}. The total number of iterations required is related to the Kullback--Leibler divergence between the prior and the posterior. However, a strict precision criterion can cause nested sampling to continue beyond the point of maximum information gain from prior to posterior (as reflected by $\KL$). For example, in \Cref{fig:dns_plots}, the iteration count ratio for Pantheon between $w_0w_a$CDM and $\Lambda$CDM is approximately 1:2.66, whilst the corresponding $\KL$ ratio (shown in \Cref{fig:dkl_single}) is only 1:1.48. This discrepancy arises because the precision criterion requires nested sampling to carry on sampling even after the bulk of information gain has been extracted, ensuring convergence to a high-precision evidence estimate.

\textbf{Prior Volume in a Three-Phase Run.} The varying number of live points during the compression and deletion phases means the simple exponential approximation for prior volume does not hold throughout the entire run. The rigorous relationship between iteration number and prior volume is given by \Cref{eq:logX_general}, which correctly accounts for the changing live point count $n_k$ throughout all three phases of the run~\cite{Hu2023aeons}. This provides a precise mapping between the iteration number (x-axis of Figure~\ref{fig:dns_plots}) and the expected log-prior volume being explored.

Figure~\ref{fig:dns_plots} demonstrates two key features: (1) for the same dataset but different models (e.g., $\Lambda$CDM and $w_0w_a$CDM for \texttt{Pantheon}, shown in orange and green), the iteration count increases only slightly for the more complex model; (2) for the same model but different datasets (e.g., $\Lambda$CDM with \texttt{Planck+lensing} vs.\ \texttt{Pantheon}, shown in blue and orange), the iteration count varies significantly, with \texttt{Planck}'s larger parameter space requiring substantially longer run times. This behaviour is also illustrated in Figures~\ref{fig:dkl_single}, \ref{fig:dkl_combo_part1}, and \ref{fig:dkl_combo_part2}, where the KL divergence values remain similar across models (across rows) but vary greatly across datasets (across columns).

\subsection{Cosmological Datasets}
\label{ssec:cosmological_datasets}

We analyse a comprehensive set of cosmological observations spanning multiple redshift ranges and probing different physical phenomena. CMB observations probe the early universe at $z \approx 1100$, while late-universe probes including baryon acoustic oscillations, Type Ia supernovae, and weak gravitational lensing constrain the expansion history and structure formation across cosmic time. To constrain the model parameters, we perform MCMC and nested sampling runs using the Cobaya framework~\cite{Torrado2021Cobaya}, which interfaces with the PolyChord sampler~\cite{Handley2015PolychordI,Handley2015PolychordII} and the CAMB Boltzmann code~\cite{Lewis:1999bs}. A full list of the likelihood packages used for this analysis is provided in Table~\ref{tab:datasets}.

\begin{table*}
\centering
\begin{tabular}{p{7cm}p{8cm}}
\hline\hline
\textbf{Dataset} & \textbf{Likelihood} \\
\hline\hline
\multicolumn{2}{l}{\textbf{Cosmic Microwave Background}} \\
\hline
Planck~\cite{Planck2020likelihoods} & \texttt{planck\_2018\_lowl.TT} \\
& \texttt{planck\_2018\_lowl.EE} \\
& \texttt{planck\_2018\_highl\_plik.TTTEEE} \\
& \texttt{planck\_2018\_highl\_plik.SZ} \\[0.3ex]
Planck with CMB lensing~\cite{Planck2020likelihoods,Planck2020lensing} & \texttt{planck\_2018\_lowl.TT} \\
& \texttt{planck\_2018\_lowl.EE} \\
& \texttt{planck\_2018\_highl\_plik.TTTEEE} \\
& \texttt{planck\_2018\_highl\_plik.SZ} \\
& \texttt{planck\_2018\_lensing.clik} \\[0.3ex]
CamSpec~\cite{Planck2020likelihoods,CamSpec2020} & \texttt{planck\_2018\_lowl.TT} \\
& \texttt{planck\_2018\_lowl.EE} \\
& \texttt{planck\_2018\_highl\_CamSpec2021.TTTEEE} \\[0.3ex]
CamSpec with CMB lensing~\cite{Planck2020likelihoods,CamSpec2020,Planck2020lensing} & \texttt{planck\_2018\_lowl.TT} \\
& \texttt{planck\_2018\_lowl.EE} \\
& \texttt{planck\_2018\_highl\_CamSpec2021.TTTEEE} \\
& \texttt{planck\_2018\_lensing.clik} \\[0.3ex]
CMB Lensing~\cite{Planck2020lensing} & \texttt{planck\_2018\_lensing.clik} \\[0.3ex]
BICEP~\cite{BICEP2021} & \texttt{bicep\_keck\_2018} \\
\hline
\multicolumn{2}{l}{\textbf{Baryon Acoustic Oscillations}} \\
\hline
SDSS~\cite{2012MNRAS.423.3430B,Ross2015,Alam2021} & \texttt{bao.sixdf\_2011\_bao} \\
& \texttt{bao.sdss\_dr7\_mgs} \\
& \texttt{bao.sdss\_dr16\_baoplus\_lrg} \\
& \texttt{bao.sdss\_dr16\_baoplus\_elg} \\
& \texttt{bao.sdss\_dr16\_baoplus\_qso} \\
& \texttt{bao.sdss\_dr16\_baoplus\_lyauto} \\
& \texttt{bao.sdss\_dr16\_baoplus\_lyxqso} \\
\hline
\multicolumn{2}{l}{\textbf{Type Ia Supernovae}} \\
\hline
SH$_0$ES~\cite{Riess2021,Scolnic2018} & \texttt{H0.riess2020Mb} \\
& \texttt{sn.pantheon} \\[0.3ex]
Pantheon~\cite{Scolnic2018} & \texttt{sn.pantheon} \\
\hline
\multicolumn{2}{l}{\textbf{Weak Lensing}} \\
\hline
DES~\cite{Abbott2018} & \texttt{des\_y1.joint} \\
\hline\hline
\end{tabular}
\caption{Cosmological datasets and their corresponding likelihood components used in the analysis. Datasets are grouped by observational type with references to the actual data packages and implementation repositories used. Likelihood names correspond to those used by \texttt{Cobaya}. Citations for the likelihoods were obtained using \texttt{cobaya-bib}.}
\label{tab:datasets}
\end{table*}

\subsubsection{Planck}
This dataset comprises high-precision measurements of CMB temperature and polarisation anisotropies from the surface of last scattering ($z \approx 1100$) using the \emph{Plik} high-$\ell$ likelihood~\cite{Planck2018params}. Our analysis utilises four likelihood components: low-$\ell$ temperature and E-mode polarisation (\texttt{planck\_2018\_lowl.TT}, \texttt{planck\_2018\_lowl.EE}) covering $\ell = 2$--29, the high-$\ell$ TTTEEE likelihood (\texttt{planck\_2018\_highl\_plik.TTTEEE}) spanning $\ell = 30$--2508, and a Sunyaev-Zel'dovich (SZ) foreground prior (\texttt{planck\_2018\_highl\_plik.SZ}). Temperature fluctuations arise from acoustic oscillations in the primordial photon-baryon plasma~\cite{Peebles1970,HuWhite1997}, whilst polarisation E-modes trace Thomson scattering during the recombination and reionisation epochs. This dataset provides strong constraints on the fundamental cosmological parameters: the baryon density $\Omega_b h^2$, cold dark matter density $\Omega_c h^2$, Hubble parameter $H_0$, primordial amplitude $A_s$, spectral index $n_s$, and reionisation optical depth $\tau$.

\subsubsection{Planck with CMB Lensing}
This dataset combines the Planck 2018 CMB measurements with the CMB lensing reconstruction (\texttt{planck\_2018\_lensing.clik})~\cite{Planck2018params,Planck2020lensing}. The lensing likelihood is added to the four baseline Planck components (low-$\ell$ TT and EE, high-$\ell$ TTTEEE, and SZ foreground). This combination provides enhanced constraints by breaking geometric degeneracies and improving measurements of the matter density $\Omega_m$ and the clustering amplitude $\sigma_8$ through an independent probe of large-scale structure growth.

\subsubsection{CamSpec}
This dataset represents an alternative high-$\ell$ analysis of Planck 2018 data, using the CamSpec 2021 likelihood (\texttt{planck\_2018\_highl\_CamSpec2021.TTTEEE}) combined with the same low-$\ell$ likelihoods as the baseline Planck analysis~\cite{CamSpec2020}. CamSpec employs distinct foreground modelling and power spectrum estimation compared to the official \emph{Plik} pipeline, including different approaches to dust cleaning and the treatment of systematics. It provides an independent systematic cross-check, which is particularly valuable for assessing the robustness of cosmological parameter constraints to the choice of analysis methodology.

\subsubsection{CamSpec with CMB Lensing}
This dataset combines the CamSpec CMB analysis (low-$\ell$ TT and EE plus high-$\ell$ CamSpec2021 TTTEEE) with the CMB lensing reconstruction (\texttt{planck\_2018\_lensing.clik})~\cite{CamSpec2020,Planck2020lensing}. It provides an independent systematic cross-check with enhanced parameter constraints from lensing, and is particularly valuable for assessing whether tensions in $\Omega_m$ and $\sigma_8$ persist across different CMB analysis pipelines.

\subsubsection{CMB Lensing}
This dataset is the standalone measurement of the lensing potential power spectrum, derived from the gravitational deflection of CMB photons by intervening large-scale structure~\cite{Planck2020lensing}. The lensing reconstruction (\texttt{planck\_2018\_lensing.clik}) uses quadratic estimators~\cite{HuOkamoto2002} on Planck temperature and polarisation maps to extract the lensing convergence signal. It probes the matter distribution and structure growth over cosmic history (primarily $z \sim 0.5$--5), breaking geometric degeneracies and enhancing constraints on $\Omega_m$, $\sigma_8$, and the sum of neutrino masses $\sum m_\nu$. We run CMB lensing alone to enable tension quantification analysis on the rest of the CMB data with and without lensing.

\subsubsection{BICEP}
This dataset consists of degree-scale B-mode polarisation measurements from the BICEP/Keck Array 2018 data release (\texttt{bicep\_keck\_2018}), which searches for primordial gravitational waves from inflation~\cite{BICEP2021}. Observations at 95, 150, and 220 GHz from the South Pole target the cleanest sky region with the lowest Galactic foreground contamination. B-modes can originate from tensor perturbations (inflationary gravitational waves) or from the weak gravitational lensing of E-modes, which acts as a foreground in this search. The BK18-only constraints on the tensor-to-scalar ratio are $r_{0.05} < 0.06$ (95\% CL), directly probing the inflationary energy scale through the relation $V^{1/4} \propto r^{1/4}$~\cite{Planck2018params}.

\subsubsection{SDSS}
This dataset is a compilation of baryon acoustic oscillation (BAO) measurements from seven independent surveys spanning $z = 0.1$ to $z > 2$~\cite{2012MNRAS.423.3430B,Ross2015,Alam2021}. We combine measurements from: 6dFGS (\texttt{bao.sixdf\_2011\_bao}, $z = 0.106$), SDSS DR7 MGS (\texttt{bao.sdss\_dr7\_mgs}, $z = 0.15$), and five eBOSS DR16 tracers covering $z = 0.698$--2.33 using luminous red galaxies (LRG), emission line galaxies (ELG), quasars (QSO), plus the Lyman-$\alpha$ forest auto-correlation and its cross-correlation with quasars at $z > 2$. The BAO feature represents the imprint of primordial sound waves at recombination~\cite{EisensteinHu1998,Eisenstein2005,Cole2005}, with a characteristic sound horizon scale of $r_{\rm drag} \approx 147$ Mpc for Planck-like $\Lambda$CDM cosmologies~\cite{Planck2018params}. This provides a standard ruler that measures both the angular diameter distance $D_A(z)$ and the Hubble parameter $H(z)$ as functions of redshift, thereby constraining $\Omega_m$, $H_0$, and dark energy dynamics.

\subsubsection{SH$_0$ES}
This dataset provides a Gaussian prior on the Type Ia supernova absolute magnitude $M_b$, derived from the local distance ladder~\cite{Riess2021,Scolnic2018}. We implement this through the \texttt{H0.riess2020Mb} likelihood, which sets $M_b = -19.263 \pm 0.049$ mag based on HST observations of Cepheid variable stars in SN Ia host galaxies with three geometric anchors (Milky Way parallaxes, the Large Magellanic Cloud, and the NGC 4258 water maser). This $M_b$ prior is used alongside the Pantheon SN Ia dataset (\texttt{sn.pantheon} with \texttt{use\_abs\_mag: true}) to derive an $H_0$ value of $73.2 \pm 1.3$ km s$^{-1}$ Mpc$^{-1}$, anchoring the local expansion rate. This result creates a tension of approximately $4\sigma$ with the CMB-inferred value of $H_0 \approx 67$ km s$^{-1}$ Mpc$^{-1}$ within the $\Lambda$CDM model, motivating searches for new physics or systematic effects.

\subsubsection{Pantheon}
The Pantheon sample is a compilation of 1048 spectroscopically confirmed Type Ia supernovae from Pan-STARRS1 (PS1), SDSS, SNLS, low-$z$ surveys, and HST (\texttt{sn.pantheon})~\cite{Scolnic2018}. It covers the redshift range $0.01 < z < 2.3$ with standardised peak magnitudes corrected for light-curve shape and colour using the SALT2 fitter. The luminosity distance-redshift relation $d_L(z)$ directly probes the expansion history $H(z) = H_0 E(z)$ through the integral $d_L(z) = c(1+z) \int_0^z dz'/H(z')$, providing evidence for cosmic acceleration~\cite{Riess1998,Perlmutter1999} at $z \sim 0.5$ and constraining $\Omega_m$ and the dark energy equation of state $w$. When analysed without an external $H_0$ calibration, this dataset constrains the degenerate product $H_0 \times M_b$ rather than absolute distances.

\subsubsection{DES}
This dataset is the Dark Energy Survey Year 1 (DES Y1) ``3×2pt'' analysis (\texttt{des\_y1.joint})~\cite{Abbott2018,Kilbinger2015}, which combines three two-point correlation functions: cosmic shear (the weak lensing auto-correlation of background galaxies at $z \sim 0.2$--1.3), galaxy clustering (the angular auto-correlation of foreground lens galaxies in five tomographic redshift bins), and galaxy-galaxy lensing (the cross-correlation between lens positions and source shears). This joint analysis, spanning 1321 deg$^2$, probes both the expansion history through geometric effects and structure growth through gravitational lensing. It primarily constrains the matter density $\Omega_m$ and the clustering amplitude through the parameter combination $S_8 = \sigma_8(\Omega_m/0.3)^{0.5}$. The DES Y1 3×2pt analysis finds $S_8 = 0.773 \pm 0.026$, showing a mild tension of approximately $2\sigma$ with the higher value of $S_8 \approx 0.83$ inferred from Planck.

\subsubsection{Dataset Selection and Temporal Context}
\label{sssec:dataset_temporal_context}
The datasets used in this analysis represent a comprehensive and powerful combination of cosmological probes available at the time this work was initiated. We acknowledge that some of these have since been superseded by more recent releases. We are continuing to expand the \texttt{unimpeded} database, and nested sampling chains for newer datasets, including Pantheon+, Union3, and DES Y5, have been analysed and are available in our ongoing work~\cite{2025arXiv251110631O}. Our work provides a robust and self-consistent analysis for this specific, well-documented data compilation. The results should therefore be interpreted as a snapshot based on this data, providing a benchmark against which future analyses with updated datasets can be compared.

\subsubsection{Treatment of Nuisance Parameters}
\label{sssec:treatment_unconstrained_nuisance}

In every model--dataset combination of the grid, all cosmological parameters of the model listed in~\Cref{tab:cosmological_models} are sampled freely with their stated uniform priors, regardless of which parameters a given dataset constrains. Nuisance parameters are introduced by the \texttt{Cobaya} likelihood configuration in line with the corresponding source experimental collaboration, and may carry a uniform prior, a Gaussian prior, or be held fixed at a published value. The underlying configuration files are publicly available with each published nested sampling chain via the \texttt{unimpeded} Python API, and the released chains include the marginalised samples of all nuisance parameters as additional columns, enabling direct inspection of their posteriors by the user. No analytic marginalisation, profiling, or evidence approximation is applied: \texttt{PolyChord} integrates over the full joint parameter space at substantial computational cost, powered by UKRI DiRAC grants DP192 and DP264. This work therefore provides the ground-truth Bayesian result and a benchmark against which other studies that adopt approximate methods to mitigate this computational expense can compare their results.

By the lossless-compression theorem of Bevins \textit{et al.}~\cite{2022arXiv220711457B}, under conditions (i) disjoint nuisance sets between datasets and (ii) marginal prior consistency (in our setup, separability of cosmological and nuisance priors), the nuisance integrals can be marginalised analytically into effective dataset-only likelihoods $\likelihood_A^{\mathrm{eff}}(\params_c)$ and $\likelihood_B^{\mathrm{eff}}(\params_c)$. Both conditions hold by construction in our \texttt{Cobaya} setup.

The shared cosmological prior $\prior(\params_c)$ must be identical across the single-dataset and joint analyses for the $R$ statistic to be coherently defined. A nuisance parameter of dataset $A$ then enters $\evidence_{AB}$ and $\evidence_A$ through $\likelihood_A^{\mathrm{eff}}(\params_c)$ but not $\evidence_B$, while a shared cosmological parameter appears in all three. In $R$, changes to the nuisance prior $\prior(\phi_A)$ that produce a uniform scaling of $\likelihood_A^{\mathrm{eff}}$ cancel between numerator and denominator, a condition that holds in the absence of strong degeneracy between the nuisance and a shared cosmological parameter, typically the case for the datasets in our grid. Changes to a shared cosmological prior, by contrast, do not cancel and produce the prior sensitivity of $R$ discussed in~\Cref{sssec:r_statistic}. The same disjoint structure produces analogous cancellations in $\KL$ and $d$, so the nuisance contributions largely cancel in both the shared dimensionality $d_G = d_A + d_B - d_{AB}$ used in the $\chi^2_{d_G}$ tension calibration of~\Cref{eq:tension_probability} and the information ratio $Q = \KL^A + \KL^B - \KL^{AB}$. The suspiciousness $\log S = \log R - Q$ and the calibrated tension probability $p$ and significance $\sigma$ inherit this invariance, leaving the full suite of tension statistics largely insensitive to nuisance prior choices.

\subsection{Tools}
This work utilises a modified version of \texttt{Cobaya 3.5.2}\footnote{\url{https://github.com/AdamOrmondroyd/cobaya}}~\citep{cobayaascl, Torrado2021Cobaya} for sampling and modelling framework, which interfaces likelihoods from different datasets with the Boltzmann code \texttt{CAMB 1.4.2.1}~\citep{Lewis:1999bs,Howlett:2012mh,Mead_2016}. The specific likelihoods used for each dataset are listed in~\Cref{tab:datasets}. \texttt{PolyChord 1.22.1}\footnote{\url{https://github.com/PolyChord/PolyChordLite}} was used as a nested sampling tool to explore parameter spaces and generate posterior samples with 1000 live points. Subsequent analysis, including plots and tension statistic computation, was performed using \texttt{anesthetic}\footnote{\url{https://github.com/handley-lab/anesthetic}} and \texttt{unimpeded}.

\section{\texttt{unimpeded} in Action}
\label{sec:unimpeded_action}
\texttt{unimpeded} is a Python-based tool designed to streamline access to pre-computed cosmological chains and facilitate Bayesian analyses. This section outlines its installation, available data, and basic usage.

\subsection{Installation}
\label{ssec:installation}
The Python library \texttt{unimpeded} is publicly available on GitHub. To ensure a clean installation and avoid conflicts with other packages, we highly recommend creating and activating a dedicated Python virtual environment before proceeding.
\begin{center}
\fbox{\parbox{\dimexpr\linewidth-2\fboxsep-2\fboxrule}{%
\ttfamily
\begin{tabular}{@{}l@{}}
\hspace{1em}python -m venv venv\\
\hspace{1em}source venv/bin/activate\\
\end{tabular}%
}}
\end{center}
The simplest method is to install the latest stable release from the Python Package Index (PyPI) using \texttt{pip}:

\begin{center}
\fbox{\parbox{\dimexpr\linewidth-2\fboxsep-2\fboxrule}{\ttfamily
\hspace{1em}pip install unimpeded
}}
\end{center}
Alternatively, for users interested in modifying the source code or contributing to development, an editable version can be installed directly from the GitHub repository:

\begin{center}
\fbox{\parbox{\dimexpr\linewidth-2\fboxsep-2\fboxrule}{%
\ttfamily
\begin{tabular}{@{}l@{}}
\hspace{1em}git clone https://github.com/handley-lab/unimpeded\\
\hspace{1em}cd unimpeded\\
\hspace{1em}pip install -e .\\
\end{tabular}%
}}
\end{center}
The full source code, along with further documentation and examples, is hosted at the GitHub repository: \url{https://github.com/handley-lab/unimpeded}.
\texttt{unimpeded}'s tension statistics calculator (see~\Cref{ssec:tension_calculator}), chain analysis functionality and visualisation tools (see~\Cref{ssec:sampling_anesthetic}) are designed to be used jointly with \texttt{anesthetic}~\cite{Handley2019anesthetic}, which is automatically installed as a dependency of \texttt{unimpeded}. The installation instructions provided here are current as of the time of publication. For the most up-to-date guidance, readers are encouraged to refer to the official documentation at \url{https://unimpeded.readthedocs.io/en/latest/}.

\subsection{Available Models and Datasets}
\label{ssec:available_models_datasets}
\texttt{unimpeded} provides access to a growing grid of both nested sampling chains and MCMC chains generated using \texttt{Cobaya}~\cite{Torrado2021Cobaya}, with nested sampling performed by \texttt{PolyChord}~\cite{Handley2015PolychordI,Handley2015PolychordII}. \texttt{unimpeded} currently covers 8 cosmological models, detailed in~\Cref{ssec:cosmological_models}, and their prior ranges are summarised in~\Cref{tab:cosmological_models}. \texttt{unimpeded} currently presents 10 datasets and their pairwise combinations, detailed in~\Cref{ssec:cosmological_datasets}. The likelihood(s) used by \texttt{Cobaya} for each dataset's nested sampling and MCMC runs are listed in~\Cref{tab:datasets}. These chains are stored on Zenodo in csv format and are accessible directly through the \texttt{unimpeded} API (see~\Cref{ssec:loading_chains}), or from the Zenodo website.

\begin{table}
\centering
\begin{tabular}{p{2.2cm}p{1.8cm}p{2.2cm}p{6.8cm}}
\hline\hline
\textbf{Model} & \textbf{Parameter} & \textbf{Prior range} & \textbf{Definition} \\
\hline
$\Lambda$CDM & $H_0$ & [20, 100] & Hubble constant \\
 & $\tau_{\text{reio}}$ & [0.01, 0.8] & Optical depth to reionization \\
 & $\Omega_b h^2$ & [0.005, 0.1] & Baryon density parameter \\
 & $\Omega_c h^2$ & [0.001, 0.99] & Cold dark matter density parameter \\
 & $\log(10^{10}A_s)$ & [1.61, 3.91] & Amplitude of scalar perturbations \\
 & $n_s$ & [0.8, 1.2] & Scalar spectral index \\
\hline
$\Omega_k\Lambda$CDM & $\Omega_k$ & [-0.3, 0.3] & Curvature density parameter (varying curvature) \\[0.5ex]
$w$CDM & $w$ & [-3, -0.333] & Constant dark energy equation of state \\[0.5ex]
$w_0w_a$CDM & $w_0$ & [-3, 1] & Present-day dark energy equation of state \\
 & $w_a$ & [-3, 2] & Dark energy equation of state evolution (CPL parameterisation) \\[0.5ex]
$m_\nu\Lambda$CDM & $\Sigma m_\nu$ & [0.06, 2] & Sum of neutrino masses (eV) \\[0.5ex]
$A_L\Lambda$CDM & $A_L$ & [0, 10] & Lensing amplitude parameter \\[0.5ex]
$n_{\text{run}}\Lambda$CDM & $n_{\text{run}}$ & [-1, 1] & Running of spectral index ($dn_s/d\ln k$) \\[0.5ex]
$r\Lambda$CDM & $r$ & [0, 3] & Scalar-to-tensor ratio \\
\hline
\end{tabular}
\caption{Cosmological parameters for the models analysed in this work. The baseline $\Lambda$CDM model contains six fundamental parameters, with extensions adding additional parameters to test specific physical hypotheses. Prior ranges are specified based on theoretical constraints and observational bounds.}
\label{tab:cosmological_models}
\end{table}

\subsection{Loading Chains and Information}
\label{ssec:loading_chains}
The primary interface for accessing the pre-computed results is the \texttt{DatabaseExplorer} class in \texttt{unimpeded.database}. It provides a programmatic workflow for downloading nested sampling and MCMC chains and their associated metadata. This example demonstrates the standard user workflow in \texttt{python}. The process begins by instantiating the \texttt{DatabaseExplorer}, which lists the available content through its \texttt{.models} and \texttt{.datasets} attributes. Subsequently, both nested sampling (\texttt{ns}) and MCMC (\texttt{mcmc}) chains, along with their corresponding metadata files, are downloadable for a selected model-dataset combination. The code shows an example of downloading the nested sampling (\texttt{'ns'}) chains for the $\Omega_k\Lambda$CDM model (\texttt{'klcdm}) constrained by the DES and CamSpec with CMB lensing joint-dataset (\texttt{``des\_y1.joint+planck\_2018\_CamSpec''}). Correspondence between cosmological models and datasets and their \texttt{unimpeded} input strings are provided in~\Cref{tab:unimpeded_models} and~\Cref{tab:unimpeded_datasets}, respectively. The call to \texttt{dbe.download\_samples} returns a \texttt{samples} object containing the full posterior samples and prior samples, including their parameter values and importance weights. Complementarily, \texttt{dbe.download\_info} retrieves the \texttt{info} object, which is a yaml file containing the complete run settings used by \texttt{Cobaya} and \texttt{PolyChord} for the analysis. \texttt{samples} and are immediately ready for analysis with tools like \texttt{anesthetic}~\cite{Handley2019anesthetic}.

\noindent\framebox[\linewidth][l]{\parbox{0.95\linewidth}{\ttfamily\small
from unimpeded.database import DatabaseExplorer\\
\\
\# Initialise DatabaseExplorer\\
dbe = DatabaseExplorer()\\
\\
\# Get a list of currently available models and datasets\\
models\_list = dbe.models\\
datasets\_list = dbe.datasets\\
\\
\# Choose model, dataset and sampling method\\
method = 'ns'  \# 'ns' for nested sampling, 'mcmc' for MCMC\\
model = ``klcdm'' \# from models\_list\\
dataset = ``des\_y1.joint+planck\_2018\_CamSpec'' \# from datasets\_list\\
\\
\# Download samples chain\\
samples = dbe.download\_samples(method, model, dataset)\\
\\
\# Download \texttt{Cobaya} and \texttt{PolyChord} run settings\\
info = dbe.download\_info(method, model, dataset)
}}

\begin{table}[htbp]
\centering
\begin{tabular}{p{4cm}p{3cm}}
\hline\hline
\textbf{Model} & \textbf{\texttt{unimpeded} Input} \\
\hline
$\Lambda$CDM & \texttt{``lcdm''} \\[0.3ex]
$\Omega_k\Lambda$CDM & \texttt{``klcdm''} \\[0.3ex]
$w$CDM & \texttt{``wlcdm''} \\[0.3ex]
$w_0w_a$CDM & \texttt{``walcdm''} \\[0.3ex]
$A_L\Lambda$CDM & \texttt{``Alcdm''} \\[0.3ex]
$m_\nu\Lambda$CDM & \texttt{``mlcdm''} \\[0.3ex]
$n_{\text{run}}\Lambda$CDM & \texttt{``nrunlcdm''} \\[0.3ex]
$r\Lambda$CDM & \texttt{``rlcdm''} \\
\hline
\end{tabular}
\caption{Correspondence between cosmological models described in \Cref{ssec:cosmological_models} and their \texttt{unimpeded} input strings.}
\label{tab:unimpeded_models}
\end{table}

\begin{table}[htbp]
\centering
\begin{tabular}{p{6cm}p{5cm}}
\hline\hline
\textbf{Dataset} & \textbf{\texttt{unimpeded} Input} \\
\hline
Planck & \texttt{``planck\_2018\_plik\_nolens''} \\[0.3ex]
Planck with CMB lensing & \texttt{``planck\_2018\_plik''} \\[0.3ex]
CamSpec & \texttt{``planck\_2018\_CamSpec\_nolens''} \\[0.3ex]
CamSpec with CMB lensing & \texttt{``planck\_2018\_CamSpec''} \\[0.3ex]
CMB Lensing & \texttt{``planck\_2018\_lensing''} \\[0.3ex]
BICEP & \texttt{``bicep\_keck\_2018''} \\[0.3ex]
SDSS & \texttt{``bao.sdss\_dr16''} \\[0.3ex]
SH$_0$ES & \texttt{``H0.riess2020Mb''} \\[0.3ex]
Pantheon & \texttt{``sn.pantheon''} \\[0.3ex]
DES & \texttt{``des\_y1.joint''} \\
\hline
\end{tabular}
\caption{Correspondence between cosmological datasets described in \Cref{tab:datasets} and their \texttt{unimpeded} input strings.}
\label{tab:unimpeded_datasets}
\end{table}

\subsection{Tension Statistics Calculator}
\label{ssec:tension_calculator}
To perform a tension analysis between two datasets, $\data_A$ and $\data_B$, one must first run three separate nested sampling analyses to obtain the chains for: (1) $\data_A$ alone, (2) $\data_B$ alone, and (3) the joint dataset $\data_{AB}$. These full nested sampling runs across a collection of models and datasets took months to complete on a high performance computer, but \texttt{unimpeded} enables users to access these chains in seconds on a laptop, with only 2 lines of code demonstrated in this minimal working example. Please note that this functionality requires \texttt{anesthetic} to be installed in the same Python environment as \texttt{unimpeded} (see~\Cref{ssec:installation}).

\noindent\framebox[\linewidth][l]{\parbox{0.95\linewidth}{\ttfamily\small
from unimpeded.tension import tension\_calculator\\
\\
tension\_samples = tension\_calculator(method='ns',\\
\phantom{tension\_samples = tension\_calculator(}model='lcdm',\\
\phantom{tension\_samples = tension\_calculator(}datasetA='planck\_2018\_CamSpec',\\
\phantom{tension\_samples = tension\_calculator(}datasetB='des\_y1.joint',\\
\phantom{tension\_samples = tension\_calculator(}nsamples=1000)
}}
The output of the \texttt{tension\_calculator()} is an \texttt{anesthetic.samples.Samples} data structure containing the values for the tension statistic detailed in~\Cref{ssec:tension_quant_theory}, which directly correspond to the theoretical quantities defined previously:
\begin{itemize}
    \item \textbf{R statistics}: The function calculates `logR` as $\log\evidence_{AB} - \log\evidence_A - \log\evidence_B$, matching the definition of the logarithmic $R$ statistic (\Cref{sssec:r_statistic}).
    \item \textbf{Information Ratio}: `I` is computed as $\KL^A + \KL^B - \KL^{AB}$, as defined in \Cref{sssec:information_ratio}.
    \item \textbf{Suspiciousness}: `logS` is calculated as $\langle\log\likelihood\rangle_{\posterior_{AB}} - \langle\log\likelihood\rangle_{\posterior_A} - \langle\log\likelihood\rangle_{\posterior_B}$, corresponding to the practical computational form of suspiciousness from \Cref{eq:suspiciousness_likelihood_avg}.
    \item \textbf{Bayesian Model Dimensionality}: The dimensionality of the shared parameter space, `$d_G$`, is computed as $d_A + d_B - d_{AB}$, as defined in \Cref{sssec:bayesian_model_dimensionality}.
    \item \textbf{Tension Probability and Significance}: The function uses `$d_G$` and `logS` to compute the $p$-value (`p`) and its equivalent Gaussian significance (`tension`) in units of $\sigma$, as described in \Cref{sssec:p_and_sigma}.
\end{itemize}
This automated calculation provides a consistent and reproducible method for applying the full suite of Bayesian tension metrics across the large grid of datasets and models provided by \texttt{unimpeded}.

\subsection{Analysing Chains with \texttt{anesthetic}}
\label{ssec:sampling_anesthetic}
The nested sampling chains generated by \texttt{unimpeded} are readily processed and analysed using the \texttt{anesthetic} package~\cite{Handley2019anesthetic}. This package provides both quantitative statistical measures and powerful visualisation tools. A key function is \texttt{NestedSamples.stats()}, which computes summary statistics essential for cosmological inference, including the evidence \texttt{logZ}, which is essential for model comparison (see~\Cref{ssec:model_comparison}) and \texttt{logL\_P} (the posterior-averaged log-likelihood $\langle \ln \likelihood \rangle_P$). In addition to these quantitative diagnostics, \texttt{anesthetic} can be used to generate corner plots to visualise the one- and two-dimensional marginalised distributions. This functionality is particularly useful as it can plot the distributions for both the posterior and the prior samples, allowing for a direct visual assessment of the information gain for each parameter.

\subsection{Future Functions}
\label{ssec:future_functions}
Future work on \texttt{unimpeded} will focus on expanding its capabilities in three primary directions. First, we plan to develop machine learning emulators for both likelihoods and full posterior distributions. Trained on the extensive set of nested sampling chains generated by \texttt{unimpeded}, these emulators will facilitate extremely rapid parameter estimation and model exploration. Furthermore, this infrastructure will enable the application of simulation-based inference (SBI) methodologies, which are essential for cosmological analyses where the likelihood function is intractable. Second, the pre-computed grid of cosmological background and perturbation quantities will be expanded. This crucial update will incorporate the latest astronomical data from current and future surveys, including but not limited to ACT, Pantheon+, DESI, Euclid, and LISA, ensuring that \texttt{unimpeded} remains relevant for modern cosmology. Finally, we will implement importance sampling. This feature will provide a computationally inexpensive method for re-weighting existing posterior samples to account for different model assumptions, thereby significantly accelerating the process of updating cosmological constraints.

\section{Results}
\label{sec:results}

\subsection{Public Release of Chains via \texttt{unimpeded}}
We have publically released a library of nested sampling and MCMC chains, across 8 cosmological models, 10 datasets and 31 pairwise dataset combinations, as detailed in section~\ref{ssec:available_models_datasets}. These chains are accessible via the \texttt{unimpeded} Python package and are stored on Zenodo, ensuring permanent public access and citable DOIs for specific data releases. This fulfills a key objective of our DiRAC-funded projects (DP192 and 264) by providing a community resource analogous to, but extending the capabilities of, the Planck Legacy Archive.

\subsection{Parameter Estimation}
\label{ssec:parameter_estimation}

\begin{figure}[htbp]
\centering
\includegraphics[width=\textwidth]{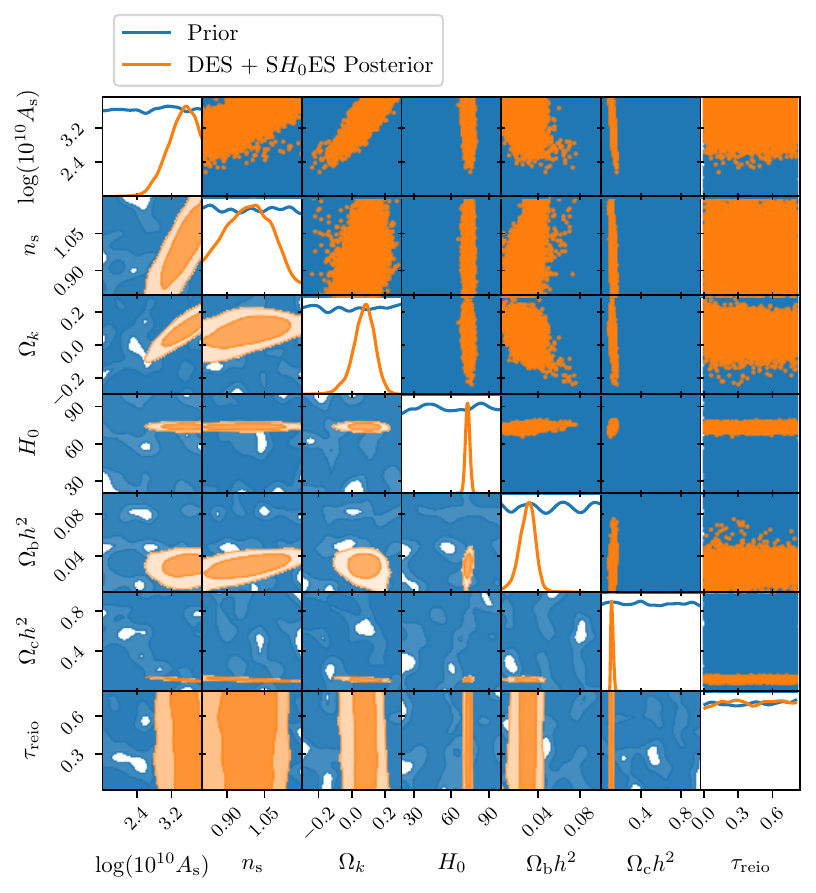}
\caption{Corner plot showing posterior distributions for the $\Omega_k\Lambda$CDM cosmological model constrained by Planck with CMB lensing + SDSS data. The diagonal panels show the one-dimensional marginalised prior (blue) and Planck with CMB lensing + SDSS posterior (orange) distributions, demonstrating the constraining power of the observational data. The lower triangular panels display the two-dimensional joint posterior and prior, where the inner (darker blue and darker orange) and outer (lighter blue and lighter orange) contours correspond to the 68\% ($1\sigma$) and 95\% ($2\sigma$) credible regions, respectively. The upper triangular panels show scatter plots of samples drawn from the posterior, visually representing parameter correlations. The posterior volume (orange) is much smaller than the prior volume (blue). This corner plot was created using \texttt{anesthetic}.}
\label{fig:prior_posterior}
\end{figure}

The \texttt{unimpeded} framework enables efficient parameter estimation by providing direct access to pre-computed full nested sampling chains and MCMC chains. This allows users to bypass the computationally intensive step of generating these chains themselves, facilitating rapid and robust cosmological inference.
We demonstrate this capability by constraining the parameters of the $\Omega_k\Lambda$CDM model using a combination of SH0ES and DES Year 1 data\footnote{The specific \texttt{unimpeded} input for this dataset combination is \texttt{H0.riess2020Mb+des\_y1.joint} (see~\Cref{tab:datasets}).}. \Cref{fig:prior_posterior} illustrates how the posterior (orange) for certain parameters are significantly more constrained compared to the broad prior (blue). The diagonal plots show the marginalised posterior probability distribution for that specific parameter. For parameters well-constrained by these datasets, such as $\log(10^{10} A_\mathrm{s})$, $n_\mathrm{s}$, and $\Omega_k$, the posteriors appear as narrow peaks demonstrating substantial information gain. However, for parameters that these particular datasets do not strongly constrain, such as $\tau_\mathrm{reio}$, the posteriors remain broad and similar to the priors. This visualisation was created using \texttt{anesthetic}.

\subsection{Model Comparison}
\label{ssec:model_comparison_results}
The Bayesian evidence values computed via nested sampling from \texttt{unimpeded} form the basis for rigorous model comparison. Here, we present a systematic comparison of eight cosmological models using both individual and combined datasets, as outlined in \Cref{ssec:available_models_datasets}. Since we used uniform priors for the set of competing models, $\Prob(\model_i) = \mathrm{constant}$, the posterior probability of a model given the data $\data$, $\Prob(\model_i|\data) = \evidence_i / \sum_j \evidence_j$ (\Cref{eq:model_prob}), provides a self-contained, normalised probability distribution over the models, allowing for a direct and intuitive ranking of their relative support from the data. Since the posterior probabilities can span many orders of magnitude, we present the natural logarithm of $\log \Prob(\model_i|\data)$, where higher values (i.e., less negative) indicate stronger evidence in favour of a given model.

\begin{figure}[htbp]
    \centering
    \includegraphics[width=\textwidth]{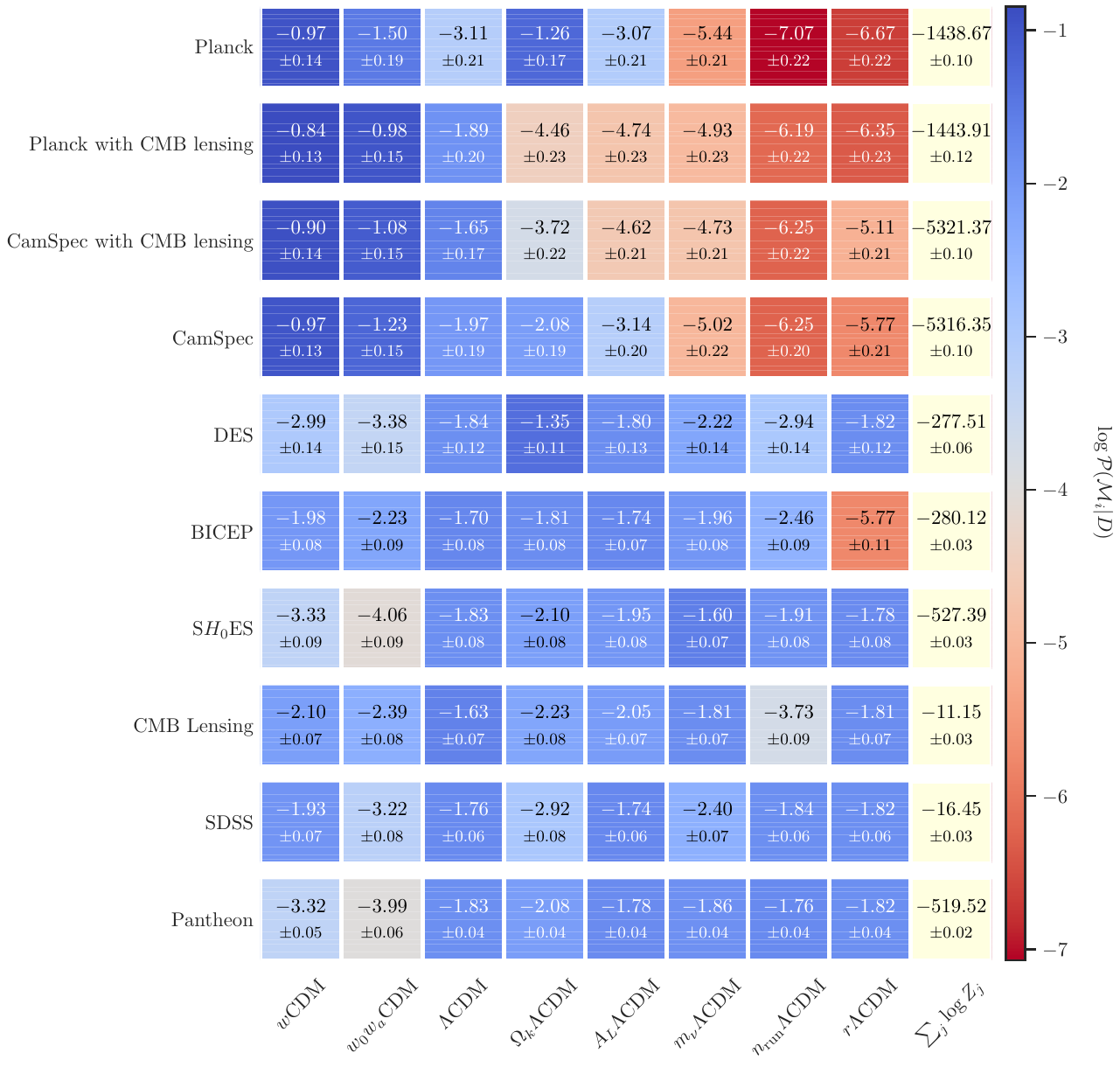}
    \caption{Heatmap of the log-posterior model probabilities, $\log \Prob(\model_i|\data)$, for each cosmological model ($x$-axis) evaluated against individual dataset ($y$-axis). Bluer colours indicate stronger statistical support for a model given the data. Comparison should only be made horizontally across models for a fixed dataset, as the sum of evidences in~\Cref{eq:model_prob} is taken over all models for that specific dataset. The final column (in yellow) shows the normalising factor $\log\left(\sum_j \evidence_j\right)$, the logarithm of the denominator of~\Cref{eq:model_prob}. The raw log-evidence for any model-dataset combination can be recovered by multiplying the probability value in that cell by the normalising factor of that row. The results show that while different datasets favour different model extensions, the base $\Lambda$CDM model emerges as the most consistently well-performing model across all individual datasets (overall blue).}
    \label{fig:model_comp_single}
\end{figure}

\begin{figure}[p]
    \vspace{-1cm}
    \centering
    \includegraphics[width=\textwidth]{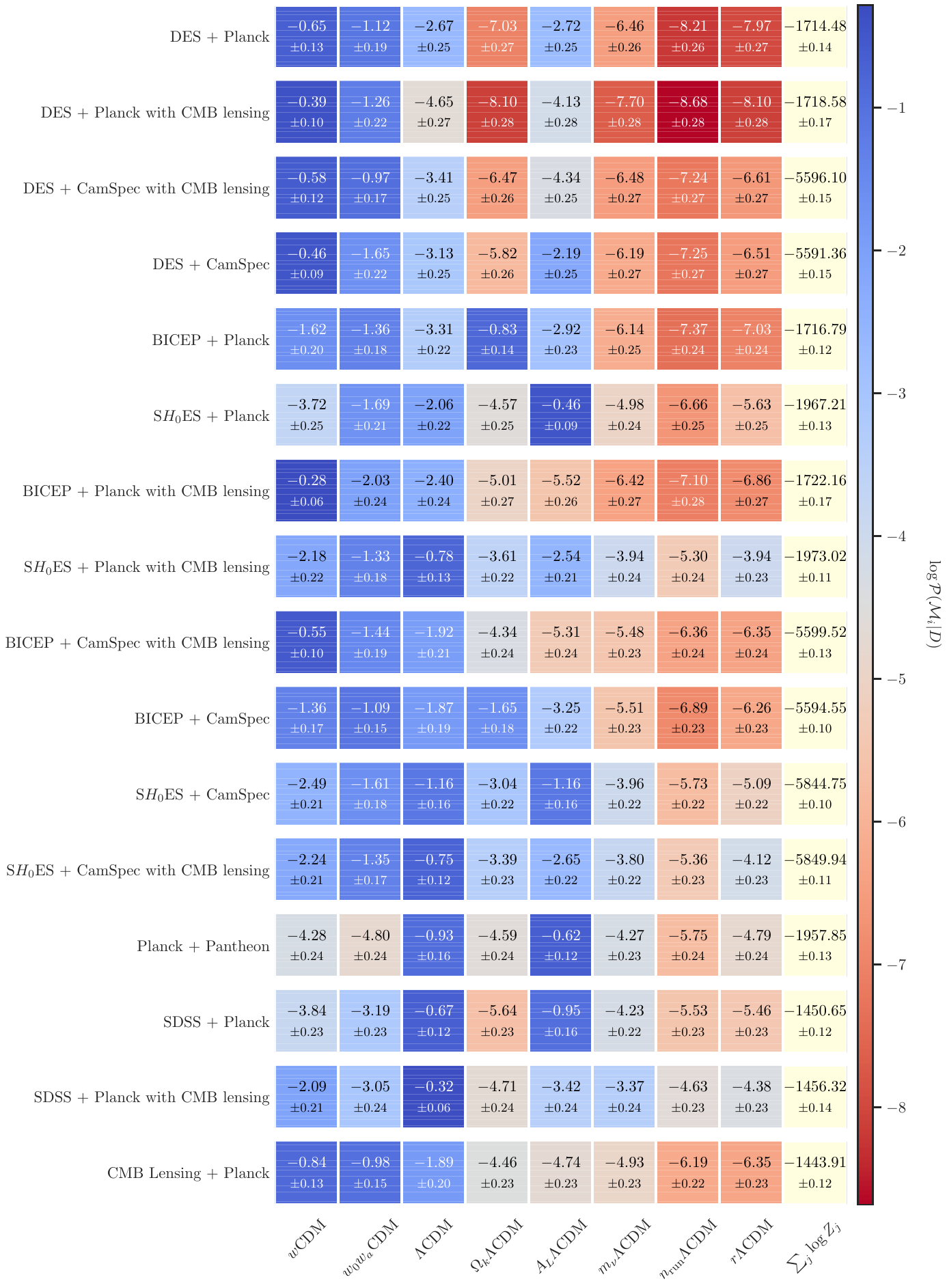}
    \caption{Same as \Cref{fig:model_comp_single}, but for combinations of datasets. The final column (in yellow) shows the normalising factor $\log\left(\sum_j \evidence_j\right)$, the logarithm of the denominator of~\Cref{eq:model_prob}, allowing recovery of raw log-evidence values by multiplying the probability in each cell by the normalising factor of that row. The combination of multiple probes sharpens the model comparison, further strengthening the preference for $\Lambda$CDM and increasing the degree to which extended models are disfavoured. Part 1 of combined datasets.}
    \label{fig:model_comp_combined_part1}
\end{figure}

\begin{figure}[p]
    \centering
    \includegraphics[width=\textwidth]{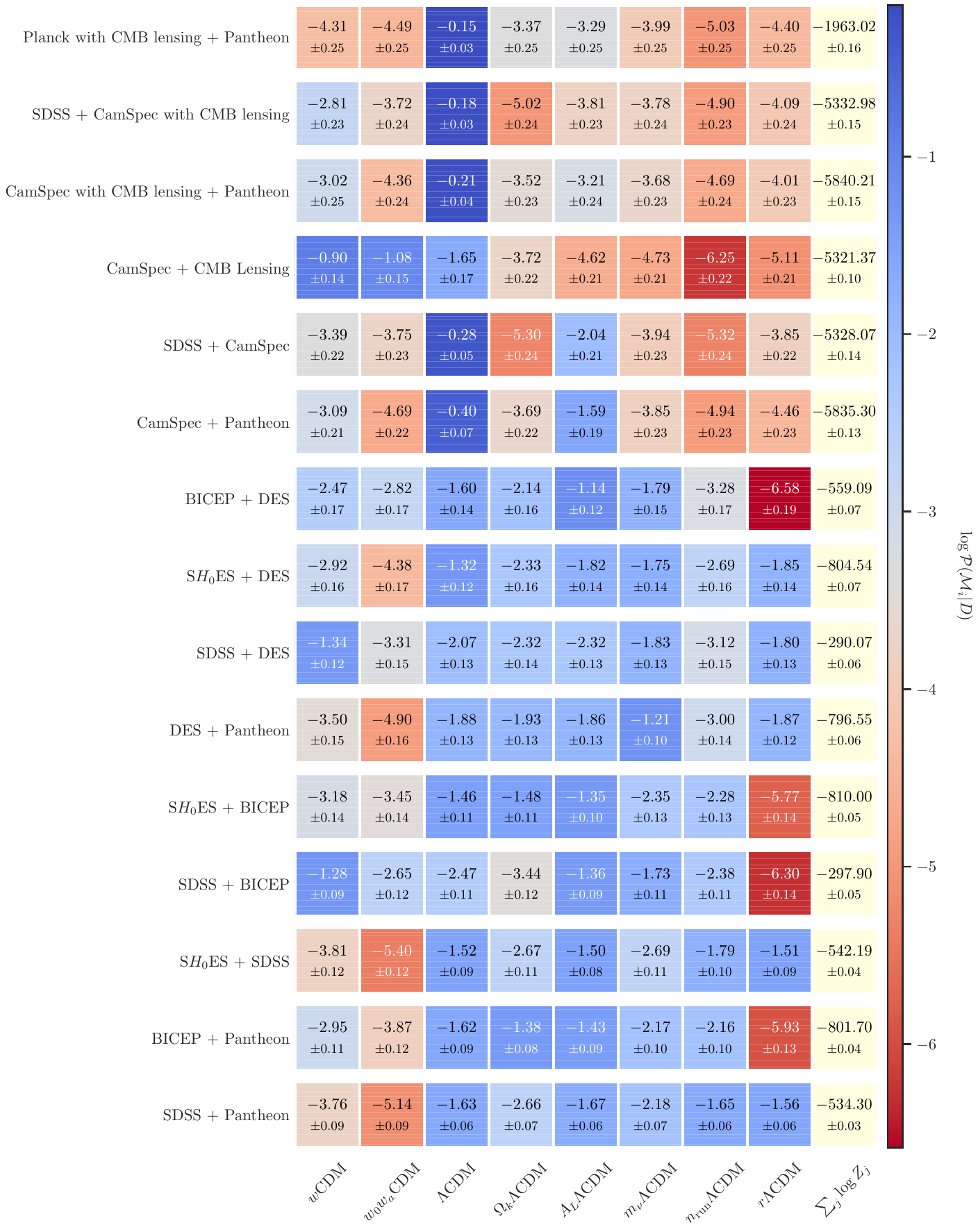}
    \caption{Same as \Cref{fig:model_comp_single}, but for combinations of datasets. The final column (in yellow) shows the normalising factor $\log\left(\sum_j \evidence_j\right)$, the logarithm of the denominator of~\Cref{eq:model_prob}, allowing recovery of raw log-evidence values by multiplying the probability in each cell by the normalising factor of that row. The combination of multiple probes sharpens the model comparison, further strengthening the preference for $\Lambda$CDM and increasing the degree to which extended models are disfavoured. Part 2 of combined datasets.}
    \label{fig:model_comp_combined_part2}
\end{figure}

The results of our model comparison are summarised in \Cref{fig:model_comp_single,fig:model_comp_combined_part1,fig:model_comp_combined_part2}. \Cref{fig:model_comp_single} presents a heatmap of $\log \Prob(\model_i|\data)$ for each of the eight models tested against 10 individual datasets. The colour scale indicates the level of support, with bluer colours corresponding to higher $\Prob(\model_i|\data)$ and redder colour indicating that a model is more disfavoured relative to the others. To structure the visualisation, the datasets ($y$-axis) are sorted by descending constraining power (model-posterior-weighted average $\langle\KL\rangle$), whilst the models ($x$-axis) are sorted by their descending $\KL$ values from the Planck with CMB lensing dataset. \Cref{fig:model_comp_combined_part1,fig:model_comp_combined_part2} show a similar analysis but for various combinations of datasets, designed to leverage their complementary constraining power.

One should note that in~\Cref{eq:model_prob}, the sum of evidences is taken over all models being compared for a specific dataset or combination of datasets. Therefore, the numerical values of $\log \Prob(\model_i|\data)$ are only comparable horizontally across models for a fixed dataset, and not vertically across datasets for a fixed model. To enable recovery of the raw log-evidence values, the final column (in yellow) of each heatmap displays the normalising factor $\log\left(\sum_j \evidence_j\right)$, which is the logarithm of the denominator of~\Cref{eq:model_prob}. The raw log-evidence for any specific model-dataset combination can be recovered by multiplying the probability value in that cell by the normalising factor of that row. 

The model preference exhibits a dependence on the specific dataset being considered. As shown in \Cref{fig:model_comp_single}, an analysis of individual datasets reveals a diversity in the preferred cosmological model. No single model is universally favoured. Instead, different probes indicate a weak preference for different extensions to the base $\Lambda$CDM model. For instance, the SDSS dataset weakly prefers the $A_L\Lambda$CDM model, whilst DES weakly prefers a non-flat universe ($\Omega_k\Lambda$CDM). The Planck primary and CamSpec datasets both weakly prefer the $w$CDM model, characterised by a constant but non-standard dark energy equation of state. Other datasets show weak preferences for a running spectral index ($n_{\mathrm{run}}\Lambda$CDM for Pantheon) or massive neutrinos ($m_\nu\Lambda$CDM for SH0ES), whilst BICEP weakly prefers the base $\Lambda$CDM model itself. Notably, the $\Lambda$CDM model, though not always exhibiting the highest $\log \Prob(\model_i|\data)$, emerges as the most consistently well-performing model amongst the eight models considered.

This picture changes when datasets are combined, as illustrated in \Cref{fig:model_comp_combined_part1,fig:model_comp_combined_part2}. In the combined analyses, the base $\Lambda$CDM model is most often the preferred scenario. We emphasise that this analysis involves comparing model performance horizontally within each dataset row; due to differences in data normalisation, a vertical comparison of log-evidence values across different datasets for a fixed model is not meaningful. 

\subsection{Constraining Power of Models and Datasets}
\label{ssec:constraining_power}
\begin{figure}[htbp]
    \centering
    \includegraphics[width=\textwidth]{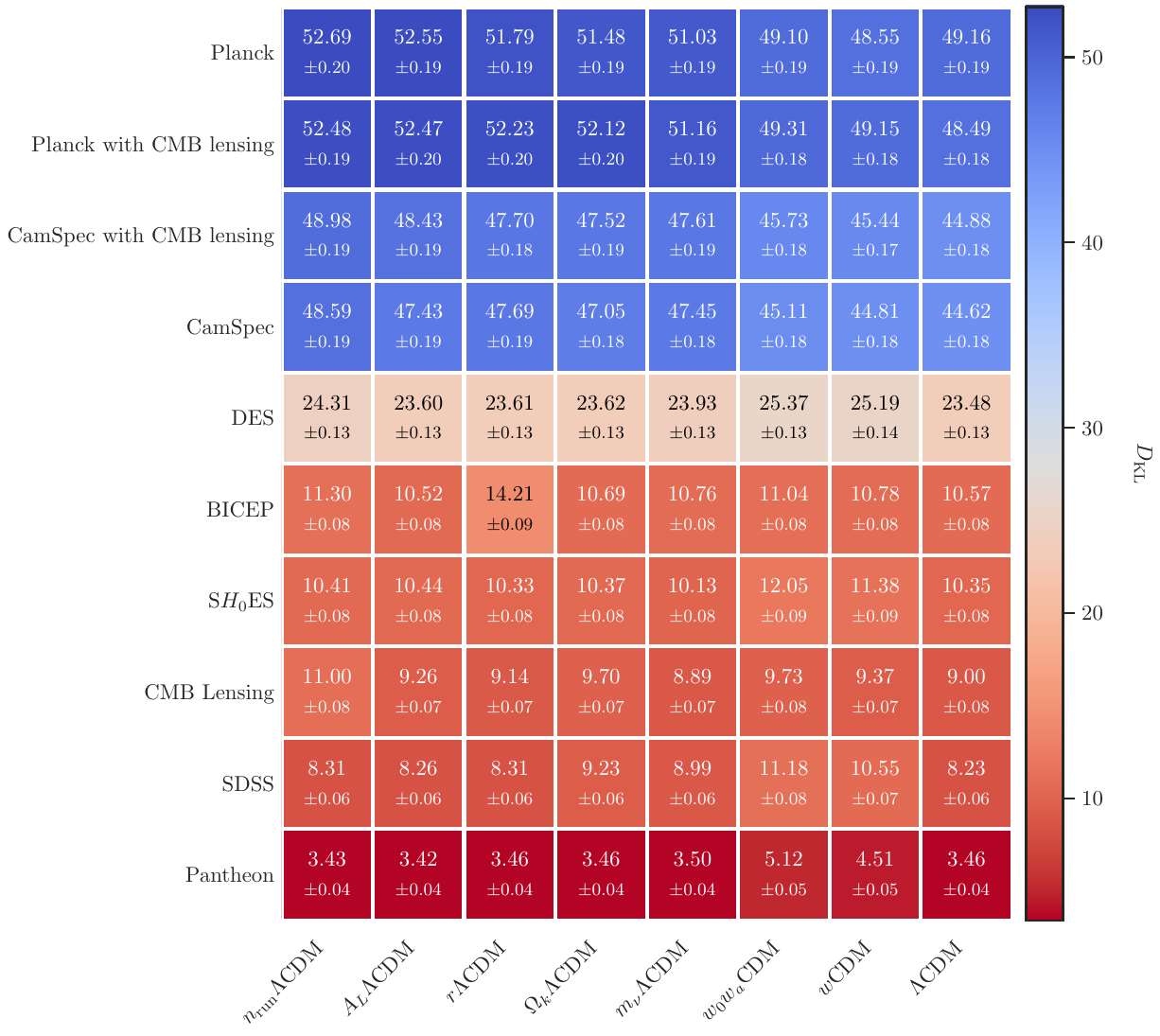}
    \caption{This heatmap illustrates the Kullback-Leibler divergence ($\KL$) for each dataset ($y$-axis) and model ($x$-axis) combination, with higher values (bluer colours) indicating a greater overall constraint. Datasets are sorted vertically by their model-posterior-weighted average $\langle \KL \rangle_{\Prob(\model)}$ (\Cref{eq:model_weighted_kl}), while models are sorted horizontally by their $\KL$ from the Planck with CMB lensing dataset. A prominent feature is the strong vertical gradient, showing that $\KL$ varies significantly among datasets but remains relatively constant across models for a given dataset. This indicates that the information gain is predominantly determined by the statistical power of the observational probe, with more constraining, information-rich datasets naturally yielding higher $\KL$ values.}
    \label{fig:dkl_single}
\end{figure}

\begin{figure}[p]
    \centering
    \includegraphics[width=\textwidth]{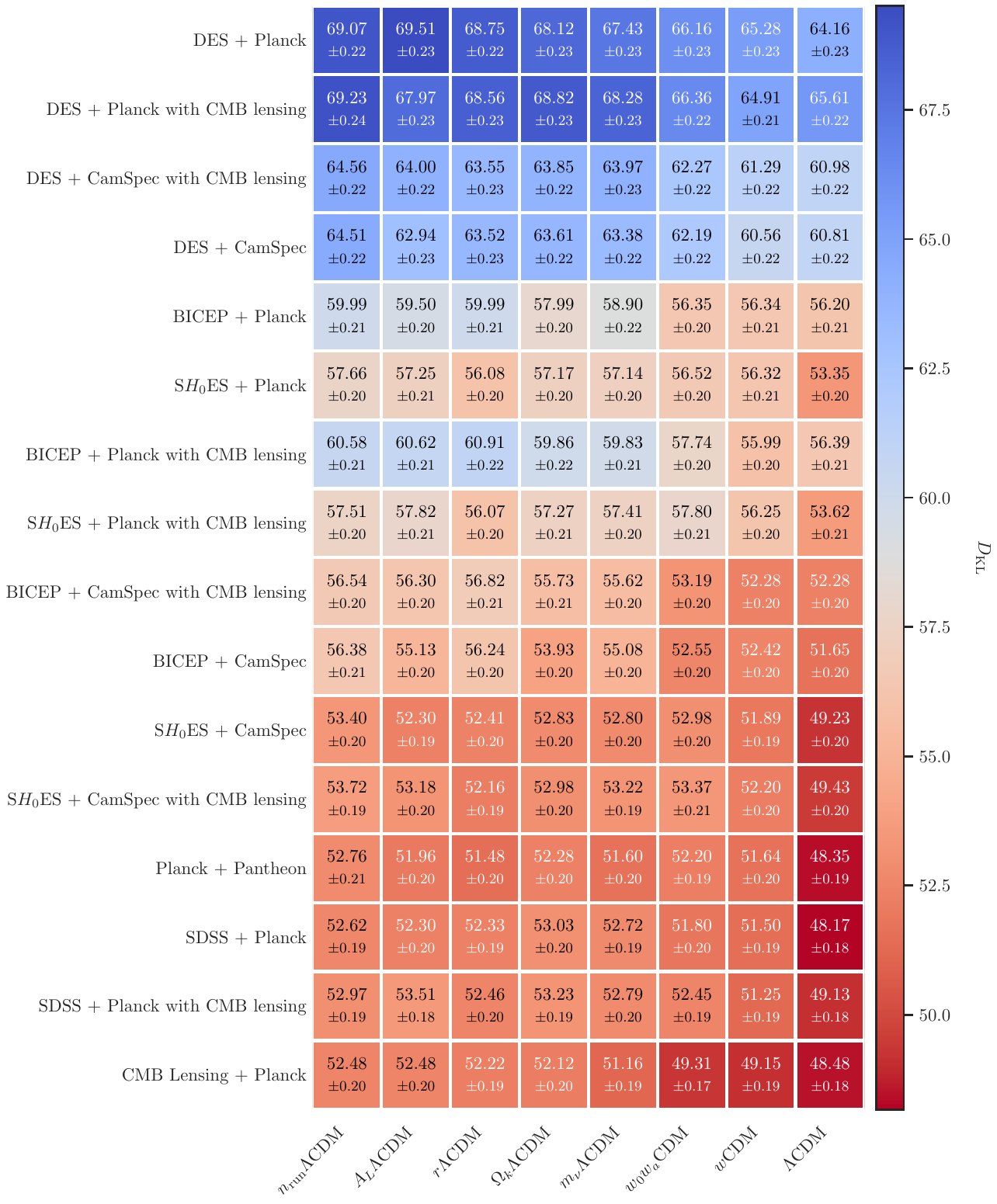}
    \caption{Same as \Cref{fig:dkl_single}, but for combinations of datasets ($y$-axis). Combined datasets yield substantially higher $\KL$ values compared to individual datasets, reflecting the enhanced constraining power from multiple complementary observational probes. The strong vertical gradient persists, with $\KL$ varying significantly among dataset combinations but remaining relatively constant across models for a given combination. Part 1 of combined datasets.}
    \label{fig:dkl_combo_part1}
\end{figure}

\begin{figure}[p]
    \centering
    \includegraphics[width=\textwidth]{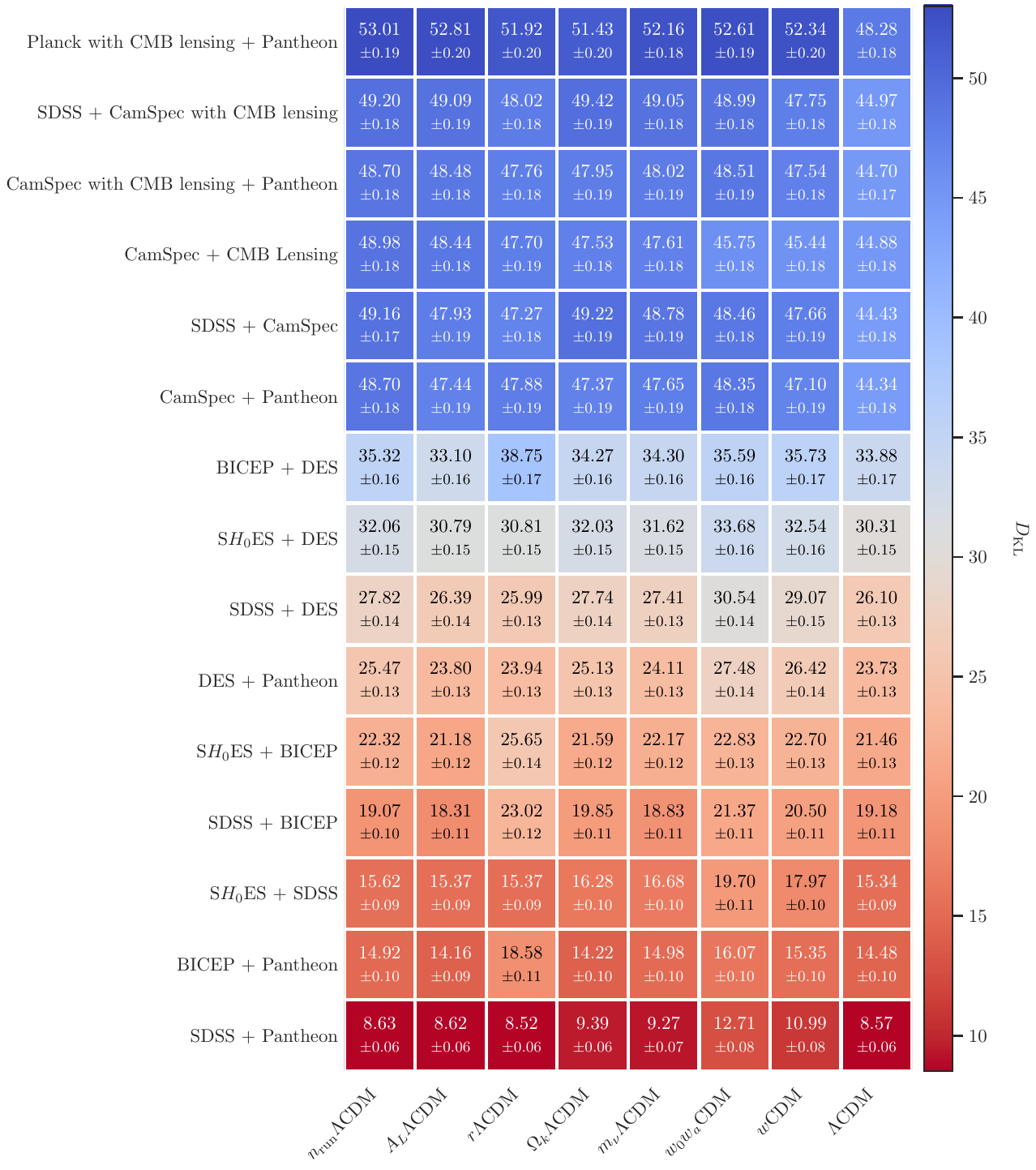}
    \caption{Same as \Cref{fig:dkl_single}, but for combinations of datasets ($y$-axis). Combined datasets yield substantially higher $\KL$ values compared to individual datasets, reflecting the enhanced constraining power from multiple complementary observational probes. The strong vertical gradient persists, with $\KL$ varying significantly among dataset combinations but remaining relatively constant across models for a given combination. Part 2 of combined datasets.}
    \label{fig:dkl_combo_part2}
\end{figure}

The Kullback-Leibler divergence $\KL$ quantifies the information gain from prior to posterior after taking into account the data, providing a measure of how much the data constrains each model-dataset combination (see~\Cref{sssec:kl_divergence} for details in theory). \Cref{fig:dkl_single} presents a heatmap of $\KL$ for each of the eight models tested against 10 individual datasets. \Cref{fig:dkl_combo_part1,fig:dkl_combo_part2} show a similar analysis but for various combinations of datasets. Higher values of $\KL$ indicate that the dataset provides stronger constraints on the model parameters, representing greater information gain from the prior to the posterior.
The datasets ($y$-axis) are sorted in descending order by the model posterior $\Prob(\model_i|\data)$-weighted average $\KL$ (\Cref{eq:model_weighted_kl}), thereby ranking each dataset by its constraining power, weighted by the posterior probability of each model. Similarly, the models along the $x$-axis are sorted in descending order according to their $\KL$ values for the Planck with CMB lensing dataset.

\subsection{Tension Quantification}
\label{ssec:tension_quantification_results}
\begin{figure}[p]
\vspace{-3cm}
\centering
\includegraphics[width=\textwidth]{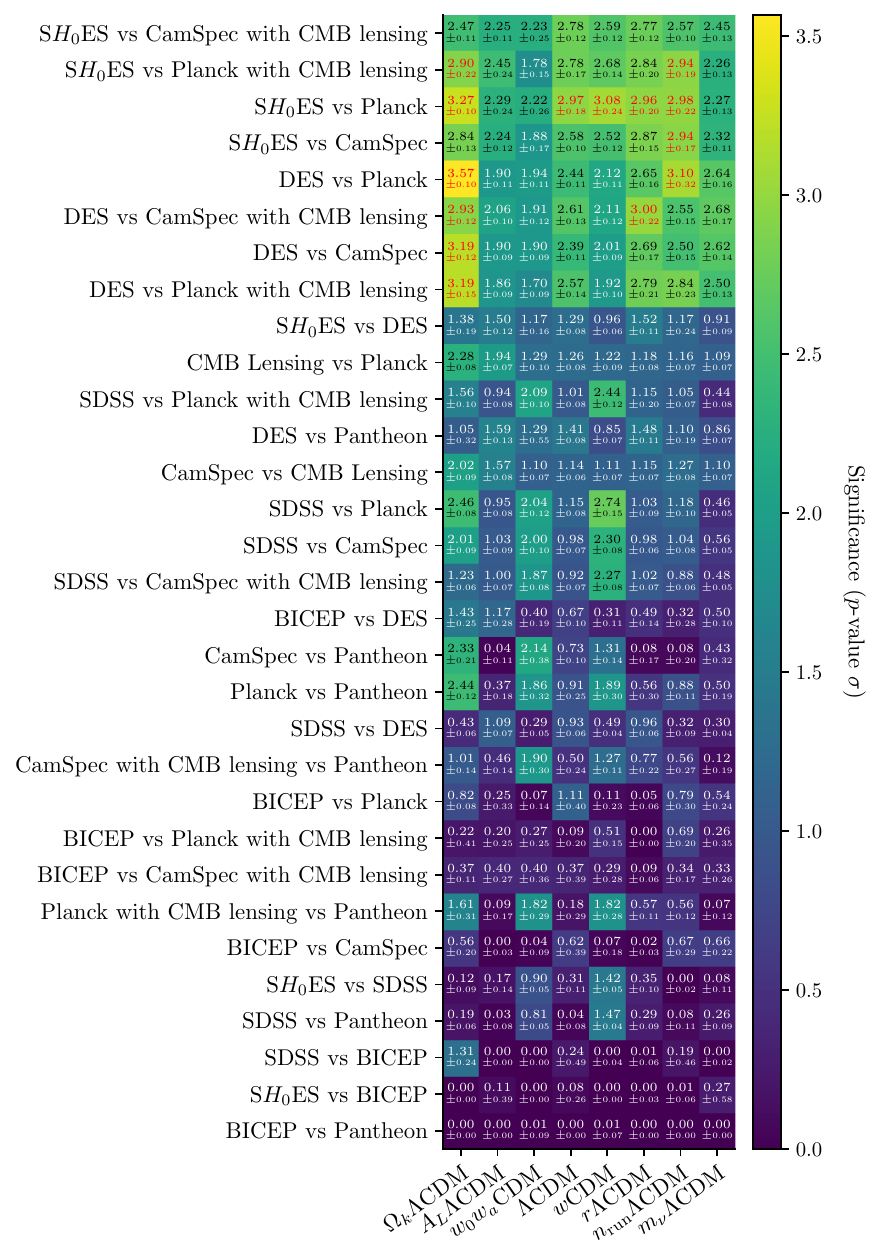}
\caption{A heatmap quantifying the tension between 31 pairwise dataset combinations ($y$-axis) across 8 cosmological models ($x$-axis). The tension is expressed as the significance in equivalent Gaussian standard deviations ($\sigma$), derived from the $p$-value, allowing for direct comparison across the grid. The dataset pairs are sorted vertically in descending order of their average tension across all models (\Cref{eq:model_weighted_sigma}), placing the most discordant combinations at the top. Values with $\sigma > 2.88$, highlighted in red, exceed the significance threshold that accounts for the look-elsewhere effect across all 248 analyses performed. This threshold is defined such that if no genuine tensions existed, only one false positive would be expected by chance (see~\Cref{sssec:look_elsewhere_effect}). We observe 14 such instances.}
\label{fig:tension_heatmap}
\end{figure}

\begin{figure}[p]
\vspace{-3cm}
\centering
\includegraphics[width=\textwidth]{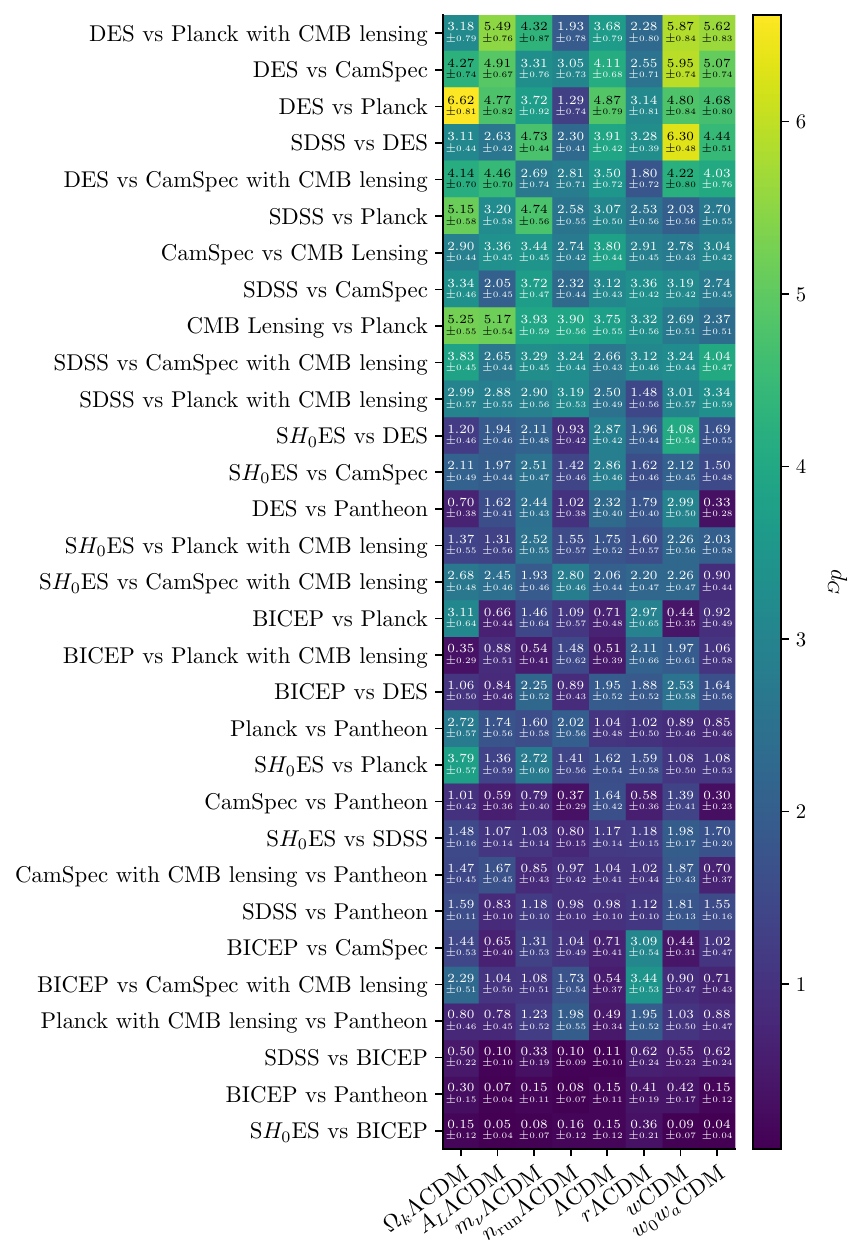}
\caption{A heatmap quantifying the Bayesian Model Dimensionality ($d_G$) for 31 pairwise dataset combinations ($y$-axis) across 8 cosmological models ($x$-axis). $d_G$ measures the effective number of constrained parameters in the shared parameter space of two datasets (see~\Cref{sssec:bayesian_model_dimensionality}), allowing for direct comparison across the grid. The dataset pairs are sorted vertically in ascending order of their average dimensionality across all models. This metric distinguishes between sharp, low-dimensional conflicts and broader, systemic disagreements.}
\label{fig:tension_d_G}
\end{figure}

\begin{figure}[p]
\vspace{-3cm}
\centering
\includegraphics[width=\textwidth]{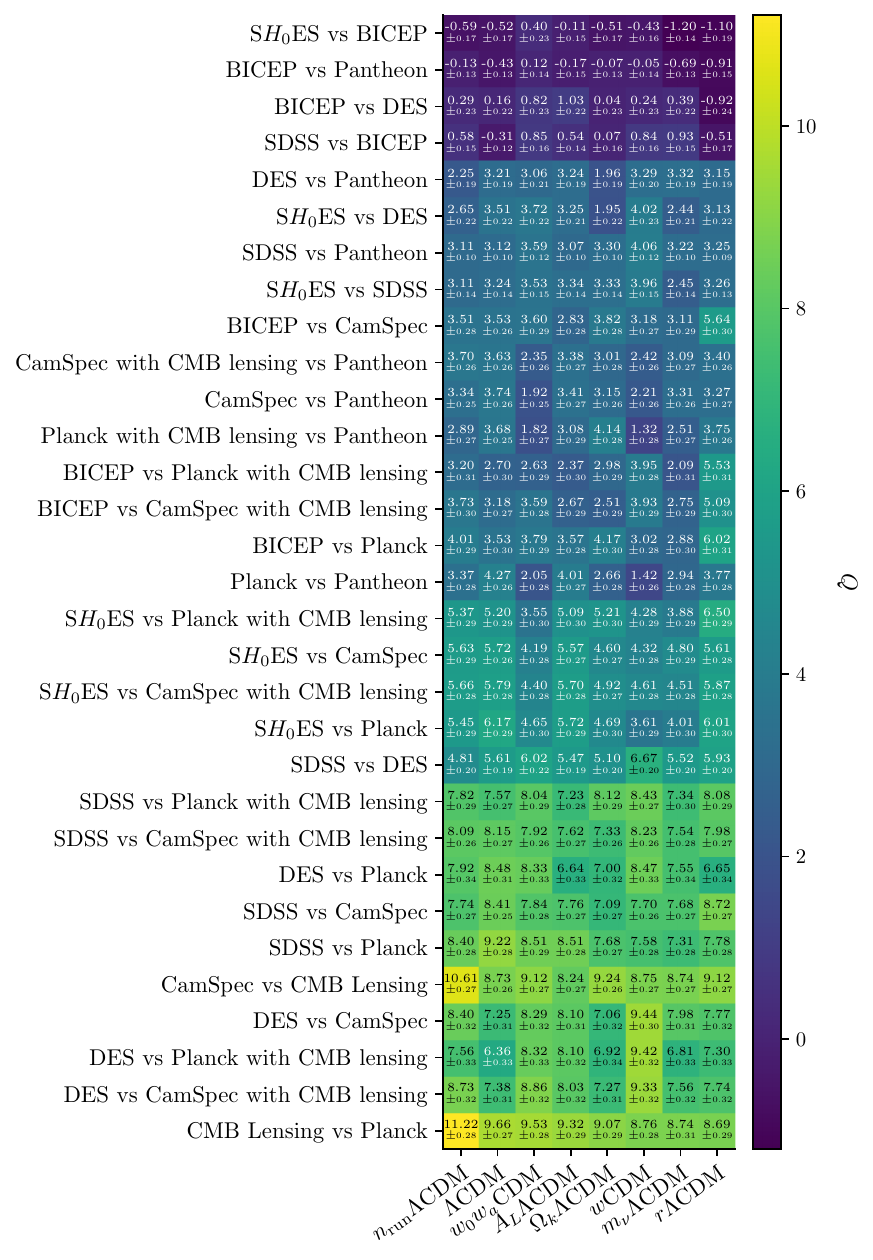}
\caption{A heatmap quantifying the tension using the Information Ratio ($Q$) for 31 pairwise dataset combinations ($y$-axis) across 8 cosmological models ($x$-axis). $Q$ quantifies tension by comparing the $\KL$ of the combined posterior relative to the individual posteriors (see~\Cref{sssec:information_ratio}), allowing for direct comparison across the grid. The dataset pairs are sorted vertically in ascending order of their average $Q$ across all models, placing the combinations with the most negative $Q$ values, and thus the strongest tension, at the top. A negative $Q$ ($Q < 0$) signifies that the volume of the combined posterior is substantially smaller than would be expected from statistically consistent datasets, pointing to minimal overlap between their individual parameter constraints. This metric therefore provides an intuitive, volume-based measure of statistical surprise.}
\label{fig:tension_I}
\end{figure}

\begin{figure}[p]
\vspace{-3.2cm}
\centering
\includegraphics[width=\textwidth]{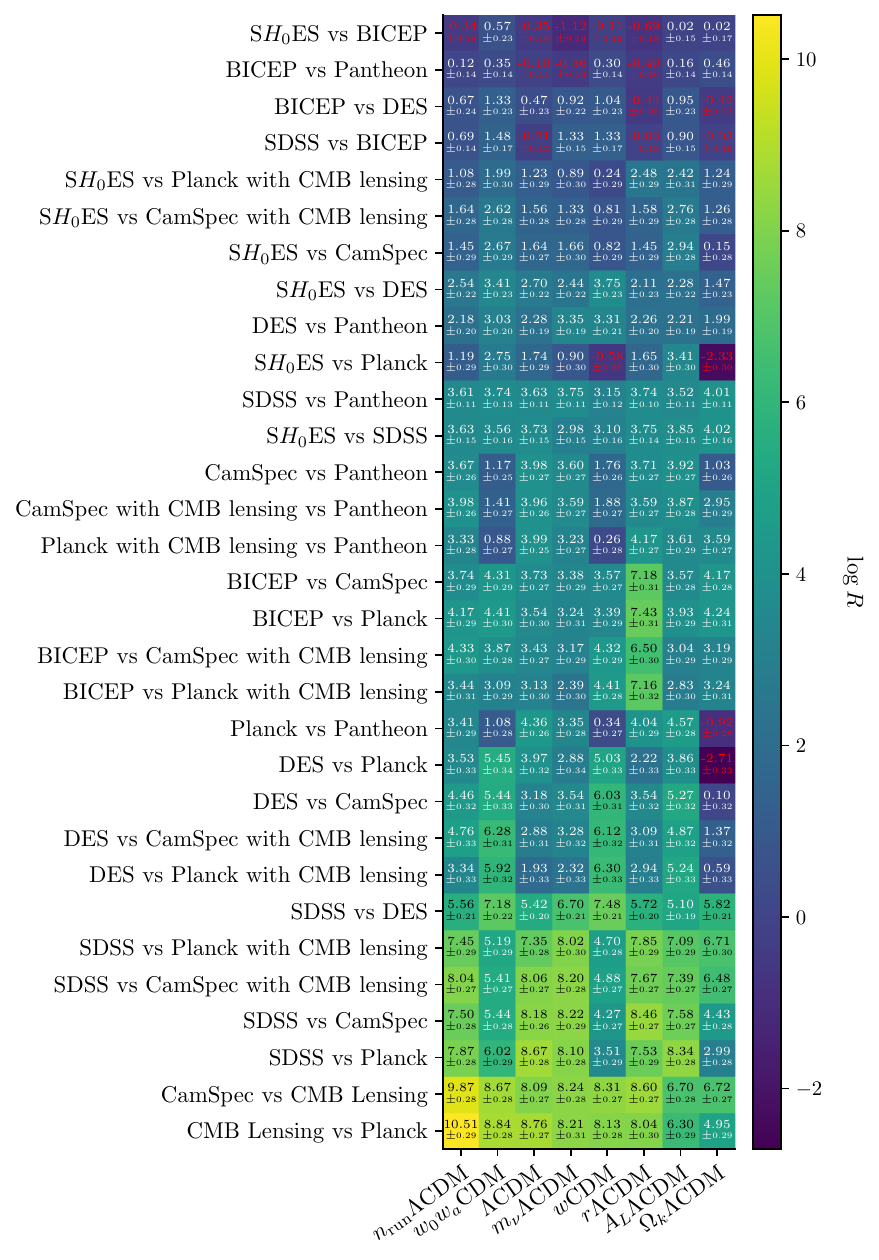}
\caption{A heatmap quantifying inter-dataset consistency using the logarithmic $R$ statistic ($\log R$) for 31 pairwise dataset combinations ($y$-axis) across 8 cosmological models ($x$-axis). The $R$ statistic is a prior-dependent measure of consistency that compares the joint Bayesian evidence of two datasets to the product of their individual evidences (see~\Cref{sssec:r_statistic}), and is interpreted relative to unity. The dataset pairs are sorted vertically in ascending order of their average $\log R$ across all models, placing the most inconsistent combinations at the top. Values of $R > 1$ ($\log R > 0$) indicate concordance, where each dataset strengthens the probability of the other. Conversely, values with $\log R < 0$ ($R < 1$), highlighted in red, indicate inconsistency, signifying that the joint probability of the data is lower than would be expected if the datasets were independent under the assumed model.}
\label{fig:tension_logR}
\end{figure}

\begin{figure}[p]
\vspace{-3cm}
\centering
\includegraphics[width=\textwidth]{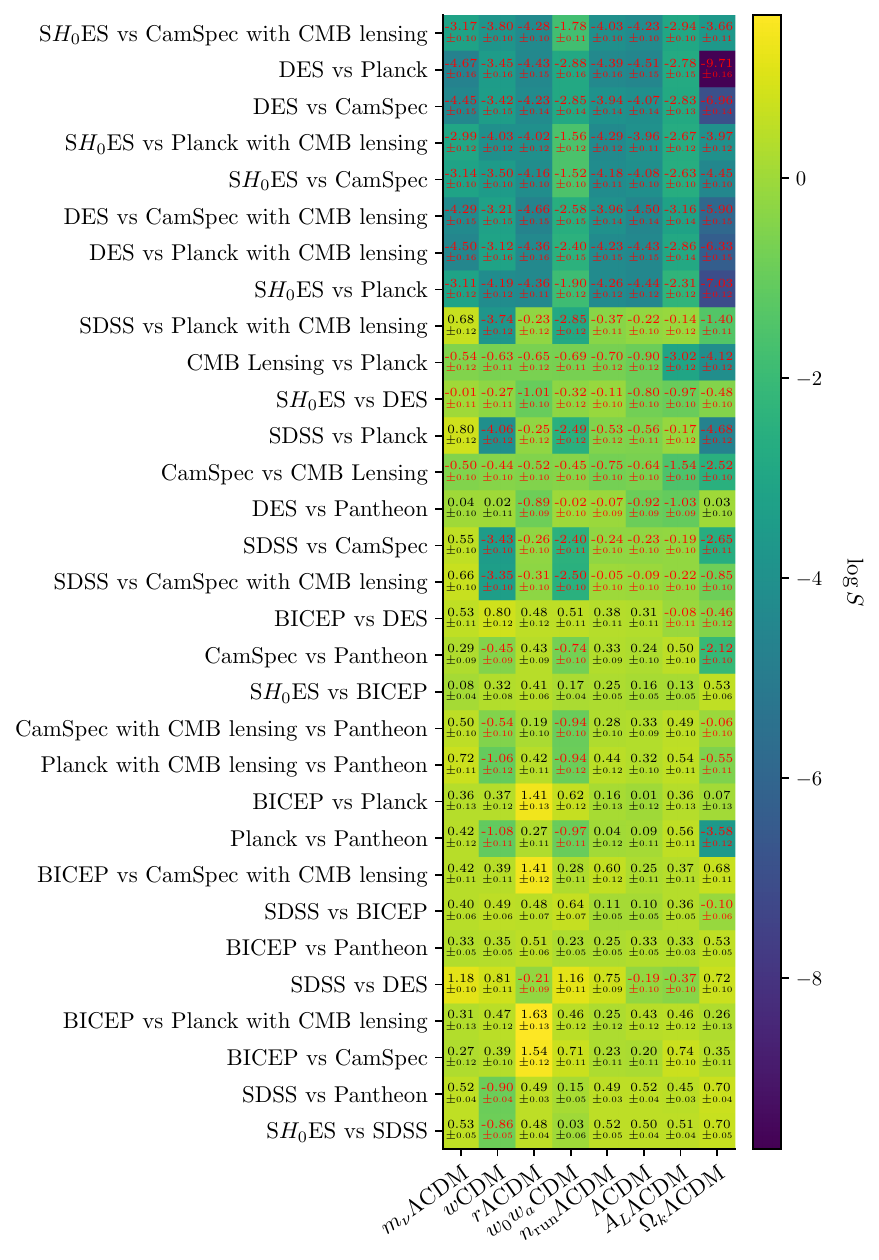}
\caption{A heatmap quantifying tension using the logarithmic Suspiciousness ($\log S$) for 31 pairwise dataset combinations ($y$-axis) across 8 cosmological models ($x$-axis). $S$ is a prior-independent metric that quantifies the statistical conflict between the likelihoods of two datasets (see~\Cref{sssec:suspiciousness}), allowing for direct comparison across the grid. The dataset pairs are sorted vertically in ascending order of their average $\log S$ (\Cref{eq:model_weighted_average}) across all models, placing the combinations with the most negative $\log S$ values, and thus the strongest tensions, at the top. Values with $\log S \ge 0$ indicate agreement, while values with $\log S < 0$, highlighted in red, indicate tension, with more negative values signifying a stronger conflict between the datasets.}
\label{fig:tension_logS}
\end{figure}

To systematically quantify the consistency between the various cosmological datasets employed in this work, we utilised the tension statistics calculator implemented in the \texttt{unimpeded} package and the nested sampling chains it offers. We performed a comprehensive tension analysis across 31 pairwise dataset combinations for each of the 8 cosmological models under consideration. For each pair, we computed the tension statistics discussed in~\cref{ssec:tension_quant_theory}, including $p$-value significance $\sigma$ (\Cref{fig:tension_heatmap}), Bayesian Model Dimensionality ($d_G$) (\Cref{fig:tension_d_G}), Information Ratio (\Cref{fig:tension_I}), $R$ statistic (\Cref{fig:tension_logR}), and Suspiciousness (\Cref{fig:tension_logS}), providing a comprehensive view of the statistical agreement between datasets. For each of the tension statistics, the dataset ($y$-axis) is ranked ascendingly or descendingly using its model posterior $\Prob(\model_i|\data)$ weighted average, as stated by \Cref{eq:model_weighted_average}.

The results of this extensive analysis are summarised in~\Cref{fig:tension_heatmap}, which presents the tension significance expressed as the equivalent Gaussian sigma ($\sigma$) of the $p$-value (see~\Cref{ssec:tension_quant_theory} for the theory and equations). A crucial feature of this representation is that the numerical $\sigma$ values are directly comparable both across rows and down columns, unlike the model comparison heatmap in~\Cref{ssec:model_comparison_results}. As discussed in~\Cref{sssec:prior_model_dependence}, these values are conditional on our chosen priors and models.

Each of the five heatmaps is sorted to bring the most concerning dataset combinations to the top, providing an immediate visual guide to potential tensions. For the $p$-value significance in~\Cref{fig:tension_heatmap}, rows are sorted in descending order of their average $\sigma$ across all models. The subsequent heatmaps for the $\log R$, $Q$, and $\log S$ statistics (\Cref{fig:tension_logR,fig:tension_I,fig:tension_logS}) are sorted in ascending order to place the most negative values—indicating strong tension—at the top. The Bayesian model dimensionality in~\Cref{fig:tension_d_G} is also sorted in ascending order. We employ red highlighting to flag values that cross specific thresholds of concern. For the $p$-value, we highlight $\sigma > 2.88$, a threshold calculated by \Cref{eq:sigma_threshold_corrected} in \Cref{sssec:look_elsewhere_effect} that accounts for the look-elsewhere effect across our 248 analyses (8 models $\times$ 31 dataset pairs). This threshold is not arbitrary, if there were no genuine tensions, we would expect exactly one result to reach $\sigma = 2.88$ purely by chance. \Cref{fig:tension_heatmap} shows 14 dataset-model combinations with $\sigma > 2.88$, significantly more than the single false positive expected under the null hypothesis. Rather than correcting individual $p$-values or $\sigma$ values (which would change as the grid expands), we apply this threshold, ensuring that $\sigma$ values remain directly interpretable independently of the grid size. For the other statistics, red flags indicate $\log R < 0$ and $\log S < 0$, signalling dataset inconsistency and direct likelihood conflict, respectively.

An analysis of the $p$-value significance in~\Cref{fig:tension_heatmap} identifies tensions that, under our specified priors and model assumptions, are consistent with the well-known tensions in cosmology. The comparisons of DES vs Planck ($\sigma=2.44{\scriptstyle\pm0.11}$ in $\Lambda$CDM) and SH0ES vs Planck ($\sigma=2.97{\scriptstyle\pm0.18}$ in $\Lambda$CDM) exhibit the highest significance, exceeding our $\sigma > 2.88$ threshold. Other comparisons involving these datasets, such as DES vs CamSpec and SH0ES vs CamSpec, also show tension in $\Lambda$CDM ($\sigma=2.39{\scriptstyle\pm0.11}$ and $\sigma=2.58{\scriptstyle\pm0.10}$, respectively). The results demonstrate model dependence; for instance, the DES vs Planck tension is alleviated in some model extensions (e.g., dropping to $\sigma=1.90{\scriptstyle\pm0.11}$ in $A_L\Lambda$CDM and $\sigma=2.12{\scriptstyle\pm0.11}$ in $w$CDM), but is exacerbated in others, reaching its maximum of $\sigma=3.57{\scriptstyle\pm0.10}$ in $\Omega_k\Lambda$CDM. The SH0ES vs Planck tension remains above $2\sigma$ in all models, with the lowest values found for $w_0w_a$CDM ($\sigma=2.22{\scriptstyle\pm0.26}$) and $m_\nu\Lambda$CDM ($\sigma=2.27{\scriptstyle\pm0.13}$), while $\Omega_k\Lambda$CDM yields the highest tension ($\sigma=3.27{\scriptstyle\pm0.10}$). This suggests that the physics introduced in certain extended models is more effective at resolving the $S_8$ tension than the Hubble tension. We note that the DES vs Planck tension is expected to relax when we extend the grid to include the DES Year 3 (Y3) data release.

The suite of five statistics provides a far more nuanced picture than the $p$-value alone, revealing crucial differences in the nature of these tensions. The rankings of the most problematic dataset pairs are broadly consistent across the $p$-value and Suspiciousness (\Cref{fig:tension_logS}) heatmaps. For example, DES vs Planck and SH0ES vs Planck comparisons populate the top rows of both, showing highly negative values for $\log S$ (e.g., for DES vs Planck in $\Lambda$CDM, $\log S = -4.51{\scriptstyle\pm0.15}$), confirming a genuine conflict between their likelihoods. These same pairs also exhibit high positive values of $Q$ in the Information Ratio heatmap (\Cref{fig:tension_I}; e.g., $Q = 8.48{\scriptstyle\pm0.31}$ for DES vs Planck in $\Lambda$CDM), reflecting the strong individual constraining power of these datasets, which places them near the bottom of the ascending-sorted $Q$ ranking rather than at the top. However, a stark disagreement emerges when comparing these to the $R$ statistic (\Cref{fig:tension_logR}). For SH0ES vs Planck, while $\log S$ is strongly negative ($-4.44{\scriptstyle\pm0.12}$ in $\Lambda$CDM), $\log R$ is positive ($+1.74{\scriptstyle\pm0.29}$), indicating concordance. This discrepancy arises because the Suspiciousness is prior-independent, whereas the $R$ statistic is not. The positive $\log R$ signifies that despite the likelihood conflict, the combined posterior is still substantially more constraining than the prior, a common feature in high-dimensional parameter spaces. This highlights the value of using the prior-independent Suspiciousness to isolate direct data conflict.

A multi-metric analysis allows a deeper physical interpretation of the tensions. The Hubble tension (SH0ES vs CMB comparisons) is characterised by high $\sigma$, high positive $Q$ and negative $\log S$, but a low Bayesian dimensionality (e.g., $d_G = 1.62{\scriptstyle\pm0.54}$ for SH0ES vs Planck in $\Lambda$CDM, see~\Cref{fig:tension_d_G}). This confirms that the conflict is sharp but concentrated in a very small number of parameter dimensions, principally $H_0$. In stark contrast, the $S_8$ tension (DES vs CMB comparisons) appears as a more systemic disagreement. For DES vs Planck in $\Lambda$CDM, not only are $\sigma$ and $\log S$ both indicative of tension, but the dimensionality is very high ($d_G = 4.87{\scriptstyle\pm0.79}$). This indicates that the datasets disagree across a wide range of parameter dimensions, representing a more fundamental inconsistency within the $\Lambda$CDM framework. The fact that this high-dimensional tension is partially resolved in certain extended models reinforces the interpretation that it may be a signature of new physics.

In summary, this comprehensive five-statistic analysis provides a detailed and robust characterisation of the consistency landscape. We find that relying on a single metric like the $p$-value can be misleading. The combined view confirms that the DES vs CMB and SH0ES vs CMB tensions are the most significant statistical conflicts in the data, but their natures are profoundly different. The $S_8$ tension is a high-dimensional problem that is partially resolved in certain model extensions, whereas the Hubble tension is a sharp, low-dimensional conflict that persists across models and is only flagged as a severe issue by prior-independent metrics like Suspiciousness. This nuanced understanding, gained by synthesising information from multiple complementary statistics, is crucial for guiding future model building and determining which dataset combinations can be reliably used for joint cosmological analyses. Caution should be exercised when combining datasets in tension. Conversely, pairs at the bottom of the rankings, such as BICEP vs Pantheon, show excellent agreement across all five metrics ($\sigma \approx 0$, $\log R \approx 0$) and can be combined with confidence.

Our findings are consistent with the curvature tension analysis of~\cite{Handley2021PRD}, which reported similar moderate tensions between Planck 2018 and CMB lensing ($\sigma = 2.49 \pm 0.07$) and between Planck 2018 and BAO ($\sigma = 3.03 \pm 0.06$) in the context of curved $\Omega_K\Lambda$CDM cosmologies. However, our model comparison results show lower Bayes factors (1.85 log units for $\Omega_k\Lambda$CDM vs $\Lambda$CDM compared to 4 log units in that work), which can be attributed to the deliberately wider priors adopted in our analysis using the Cobaya defaults. These wider priors provide greater flexibility for importance reweighting if tighter priors are desired in future analyses. Whilst that work focused specifically on the curvature parameter $\Omega_K$, our systematic analysis across eight model extensions and 31 dataset pairs provides a broader view of the tension landscape, demonstrating that the methodology is robust and the tensions persist across multiple cosmological frameworks.

\clearpage

\section{Conclusions}
\label{sec:conclusions}

In this work, we have introduced \texttt{unimpeded}, a comprehensive and publicly available resource for Bayesian cosmological analysis. We have performed a systematic nested sampling analysis of eight cosmological models, from the base $\Lambda$CDM paradigm to seven well-motivated extensions, constrained by a suite of 39 individual and combined datasets. The primary data product of this analysis is an extensive repository of MCMC and nested sampling chains, hosted on Zenodo, which we provide to the community to facilitate reproducible and extensible cosmological research. The use of deliberately wide priors ensures that these chains are a versatile resource, suitable for importance reweighting and a wide range of future studies. Future enhancements to \texttt{unimpeded} will include importance sampling capabilities, which will enable the exploration of additional cosmological models through likelihood reweighting of existing chains, circumventing the need for computationally expensive full nested sampling runs and thereby substantially broadening the range of testable models.

Our analysis yields two principal scientific conclusions. First, through a comprehensive model comparison across the eight cosmological models we consider (i.e., $\Lambda$CDM and its single-parameter and two-parameter extensions), we find that whilst individual datasets show varied preferences for model extensions, the base $\Lambda$CDM model is most frequently preferred in combined analyses, with the general trend suggesting that evidence for new physics is diluted when probes are combined. This suggests that the hints for new physics present in individual datasets do not reinforce one another in a joint analysis, at least for the simple extensions studied here. It is crucial to emphasize that this conclusion is contingent on our choice of models. More complex scenarios, such as interacting dark energy or modified gravity models with multiple parameters, were not explored. Second, by employing five complementary tension statistics, we systematically quantified the discordances between key datasets. We find the most significant tensions to be between SH0ES and Planck ($\sigma=2.97{\scriptstyle\pm0.18}$) and between DES(Y1) and Planck ($\sigma=2.44{\scriptstyle\pm0.11}$), within $\Lambda$CDM. Our multi-metric approach reveals that these tensions have profoundly different natures: the $S_8$ tension between DES and Planck is a high-dimensional disagreement ($d_G=4.87{\scriptstyle\pm0.79}$) that is mildly alleviated in models with a varying dark energy equation of state, whereas the Hubble tension between SH0ES and Planck is a sharp, low-dimensional conflict ($d_G=1.62{\scriptstyle\pm0.54}$) that persists across almost all model extensions considered.

The \texttt{unimpeded} resource provides a powerful platform for future investigations. The upgrade to DES Year 3 data is expected to clarify the status of the $S_8$ tension, and our framework provides the ideal foundation for a rapid and consistent analysis of this and other forthcoming datasets. Caution should be exercised when combining datasets in tension. By providing a standardised and accessible suite of Bayesian analysis products, we hope to accelerate progress in understanding the remaining tensions within the cosmological landscape and to robustly test the limits of the $\Lambda$CDM model.

\appendix

\acknowledgments
This work was performed using the Cambridge Service for Data Driven Discovery (CSD3), part of which is operated by the University of Cambridge Research Computing on behalf of the STFC DiRAC HPC Facility (www.dirac.ac.uk). The DiRAC component of CSD3 was funded by BEIS capital funding via STFC capital grants ST/P002307/1 and ST/R002452/1 and STFC operations grant ST/R00689X/1. DiRAC is part of the National e-Infrastructure.
W.H. acknowledges support from a Royal Society University Research Fellowship.

\bibliographystyle{JHEP}
\bibliography{biblio} 

\providecommand{\href}[2]{#2}\begingroup\raggedright\begin{thebibliography}{10}

\bibitem{Verde2019NatAs}
L.~{Verde}, T.~{Treu} and A.G.~{Riess}, \emph{{Tensions between the early and late Universe}}, \href{https://doi.org/10.1038/s41550-019-0902-0}{\emph{Nature Astronomy} {\bfseries 3} (2019) 891} [\href{https://arxiv.org/abs/1907.10625}{{\ttfamily 1907.10625}}].

\bibitem{Joudaki2017MNRAS}
S.~{Joudaki}, A.~{Mead}, C.~{Blake}, A.~{Choi}, J.~{de Jong}, T.~{Erben} et~al., \emph{{KiDS-450: testing extensions to the standard cosmological model}}, \href{https://doi.org/10.1093/mnras/stx998}{\emph{\mnras} {\bfseries 471} (2017) 1259} [\href{https://arxiv.org/abs/1610.04606}{{\ttfamily 1610.04606}}].

\bibitem{Handley2021PRD}
W.~{Handley}, \emph{{Curvature tension: Evidence for a closed universe}}, \href{https://doi.org/10.1103/PhysRevD.103.L041301}{\emph{\prd} {\bfseries 103} (2021) L041301} [\href{https://arxiv.org/abs/1908.09139}{{\ttfamily 1908.09139}}].

\bibitem{DiValentino2020NatAs}
E.~{Di Valentino}, A.~{Melchiorri} and J.~{Silk}, \emph{{Planck evidence for a closed Universe and a possible crisis for cosmology}}, \href{https://doi.org/10.1038/s41550-019-0906-9}{\emph{Nature Astronomy} {\bfseries 4} (2020) 196} [\href{https://arxiv.org/abs/1911.02087}{{\ttfamily 1911.02087}}].

\bibitem{Planck2018params}
{Planck Collaboration}, N.~{Aghanim}, Y.~{Akrami}, M.~{Ashdown}, J.~{Aumont}, C.~{Baccigalupi} et~al., \emph{{Planck 2018 results. VI. Cosmological parameters}}, \href{https://doi.org/10.1051/0004-6361/201833910}{\emph{\aanda} {\bfseries 641} (2020) A6} [\href{https://arxiv.org/abs/1807.06209}{{\ttfamily 1807.06209}}].

\bibitem{Skilling2006}
J.~Skilling, \emph{Nested sampling for general bayesian computation}, \href{https://doi.org/10.1214/06-BA127}{\emph{Bayesian Analysis} {\bfseries 1} (2006) 833}.

\bibitem{Lemos2021}
W.~{Handley} and P.~{Lemos}, \emph{{Quantifying the global parameter tensions between ACT, SPT, and Planck}}, \href{https://doi.org/10.1103/PhysRevD.103.063529}{\emph{\prd} {\bfseries 103} (2021) 063529} [\href{https://arxiv.org/abs/2007.08496}{{\ttfamily 2007.08496}}].

\bibitem{Handley2019}
W.~{Handley} and P.~{Lemos}, \emph{{Quantifying tensions in cosmological parameters: Interpreting the DES evidence ratio}}, \href{https://doi.org/10.1103/PhysRevD.100.043504}{\emph{\prd} {\bfseries 100} (2019) 043504} [\href{https://arxiv.org/abs/1902.04029}{{\ttfamily 1902.04029}}].

\bibitem{Lemos2020}
P.~{Lemos}, F.~{K{\"o}hlinger}, W.~{Handley}, B.~{Joachimi}, L.~{Whiteway} and O.~{Lahav}, \emph{{Quantifying Suspiciousness within correlated data sets}}, \href{https://doi.org/10.1093/mnras/staa1836}{\emph{\mnras} {\bfseries 496} (2020) 4647} [\href{https://arxiv.org/abs/1910.07820}{{\ttfamily 1910.07820}}].

\bibitem{2025arXiv251105470O}
D.D.Y.~{Ong} and W.~{Handley}, \emph{{$\texttt{unimpeded}$: A Public Nested Sampling Database for Bayesian Cosmology}}, \href{https://doi.org/10.48550/arXiv.2511.05470}{\emph{arXiv e-prints} (2025) arXiv:2511.05470} [\href{https://arxiv.org/abs/2511.05470}{{\ttfamily 2511.05470}}].

\bibitem{2025arXiv251110631O}
D.D.Y.~{Ong}, D.~{Yallup} and W.~{Handley}, \emph{{A Bayesian Perspective on Evidence for Evolving Dark Energy}}, \href{https://doi.org/10.48550/arXiv.2511.10631}{\emph{arXiv e-prints} (2025) arXiv:2511.10631} [\href{https://arxiv.org/abs/2511.10631}{{\ttfamily 2511.10631}}].

\bibitem{2026arXiv260305472O}
D.D.Y.~{Ong}, D.~{Yallup} and W.~{Handley}, \emph{{The Bayesian view of DESI DR2 with unimpeded: Evidence and tension in a combined analysis with CMB and supernovae across cosmological models}}, \href{https://doi.org/10.48550/arXiv.2603.05472}{\emph{arXiv e-prints} (2026) arXiv:2603.05472} [\href{https://arxiv.org/abs/2603.05472}{{\ttfamily 2603.05472}}].

\bibitem{2008ConPh..49...71T}
R.~{Trotta}, \emph{{Bayes in the sky: Bayesian inference and model selection in cosmology}}, \href{https://doi.org/10.1080/00107510802066753}{\emph{Contemporary Physics} {\bfseries 49} (2008) 71} [\href{https://arxiv.org/abs/0803.4089}{{\ttfamily 0803.4089}}].

\bibitem{kullback1951information}
S.~Kullback and R.A.~Leibler, \emph{On information and sufficiency}, \href{https://doi.org/10.1214/aoms/1177729694}{\emph{The Annals of Mathematical Statistics} {\bfseries 22} (1951) 79}.

\bibitem{2014PhRvD..90b3533S}
S.~{Seehars}, A.~{Amara}, A.~{Refregier}, A.~{Paranjape} and J.~{Akeret}, \emph{{Information gains from cosmic microwave background experiments}}, \href{https://doi.org/10.1103/PhysRevD.90.023533}{\emph{\prd} {\bfseries 90} (2014) 023533} [\href{https://arxiv.org/abs/1402.3593}{{\ttfamily 1402.3593}}].

\bibitem{2019JCAP...01..011N}
A.~{Nicola}, A.~{Amara} and A.~{Refregier}, \emph{{Consistency tests in cosmology using relative entropy}}, \href{https://doi.org/10.1088/1475-7516/2019/01/011}{\emph{\jcap} {\bfseries 2019} (2019) 011} [\href{https://arxiv.org/abs/1809.07333}{{\ttfamily 1809.07333}}].

\bibitem{2004PhRvL..92n1302H}
A.~{Hosoya}, T.~{Buchert} and M.~{Morita}, \emph{{Information Entropy in Cosmology}}, \href{https://doi.org/10.1103/PhysRevLett.92.141302}{\emph{\prl} {\bfseries 92} (2004) 141302} [\href{https://arxiv.org/abs/gr-qc/0402076}{{\ttfamily gr-qc/0402076}}].

\bibitem{2013PDU.....2..166V}
L.~{Verde}, P.~{Protopapas} and R.~{Jimenez}, \emph{{Planck and the local Universe: Quantifying the tension}}, \href{https://doi.org/10.1016/j.dark.2013.09.002}{\emph{Physics of the Dark Universe} {\bfseries 2} (2013) 166} [\href{https://arxiv.org/abs/1306.6766}{{\ttfamily 1306.6766}}].

\bibitem{2016PhRvD..93j3507S}
S.~{Seehars}, S.~{Grandis}, A.~{Amara} and A.~{Refregier}, \emph{{Quantifying concordance in cosmology}}, \href{https://doi.org/10.1103/PhysRevD.93.103507}{\emph{\prd} {\bfseries 93} (2016) 103507} [\href{https://arxiv.org/abs/1510.08483}{{\ttfamily 1510.08483}}].

\bibitem{2016JCAP...05..034G}
S.~{Grandis}, S.~{Seehars}, A.~{Refregier}, A.~{Amara} and A.~{Nicola}, \emph{{Information gains from cosmological probes}}, \href{https://doi.org/10.1088/1475-7516/2016/05/034}{\emph{\jcap} {\bfseries 2016} (2016) 034} [\href{https://arxiv.org/abs/1510.06422}{{\ttfamily 1510.06422}}].

\bibitem{2016arXiv160606273R}
M.~{Raveri}, M.~{Martinelli}, G.~{Zhao} and Y.~{Wang}, \emph{{Information Gain in Cosmology: From the Discovery of Expansion to Future Surveys}}, \href{https://doi.org/10.1103/PhysRevD.95.103502}{\emph{\prd} {\bfseries 95} (2017) 103502} [\href{https://arxiv.org/abs/1606.06273}{{\ttfamily 1606.06273}}].

\bibitem{2016MNRAS.455.2461H}
S.~{Hee}, W.J.~{Handley}, M.P.~{Hobson} and A.N.~{Lasenby}, \emph{{Bayesian model selection without evidences: application to the dark energy equation-of-state}}, \href{https://doi.org/10.1093/mnras/stv2217}{\emph{\mnras} {\bfseries 455} (2016) 2461} [\href{https://arxiv.org/abs/1506.09024}{{\ttfamily 1506.09024}}].

\bibitem{2016MNRAS.463.1416G}
S.~{Grandis}, D.~{Rapetti}, A.~{Saro}, J.J.~{Mohr} and J.P.~{Dietrich}, \emph{{Quantifying tensions between CMB and distance data sets in models with free curvature or lensing amplitude}}, \href{https://doi.org/10.1093/mnras/stw2028}{\emph{\mnras} {\bfseries 463} (2016) 1416} [\href{https://arxiv.org/abs/1604.06463}{{\ttfamily 1604.06463}}].

\bibitem{2017NatAs...1..627Z}
G.-B.~{Zhao}, M.~{Raveri}, L.~{Pogosian}, Y.~{Wang}, R.G.~{Crittenden}, W.J.~{Handley} et~al., \emph{{Dynamical dark energy in light of the latest observations}}, \href{https://doi.org/10.1038/s41550-017-0216-z}{\emph{Nature Astronomy} {\bfseries 1} (2017) 627} [\href{https://arxiv.org/abs/1701.08165}{{\ttfamily 1701.08165}}].

\bibitem{2017JCAP...10..045N}
A.~{Nicola}, A.~{Amara} and A.~{Refregier}, \emph{{Integrated cosmological probes: concordance quantified}}, \href{https://doi.org/10.1088/1475-7516/2017/10/045}{\emph{\jcap} {\bfseries 2017} (2017) 045} [\href{https://arxiv.org/abs/1706.06593}{{\ttfamily 1706.06593}}].

\bibitem{Hergt2021Bayesian}
L.T.~{Hergt}, W.J.~{Handley}, M.P.~{Hobson} and A.N.~{Lasenby}, \emph{{Bayesian evidence for the tensor-to-scalar ratio r and neutrino masses m$_{{\ensuremath{\nu}}}$ : Effects of uniform versus logarithmic priors}}, \href{https://doi.org/10.1103/PhysRevD.103.123511}{\emph{\prd} {\bfseries 103} (2021) 123511} [\href{https://arxiv.org/abs/2102.11511}{{\ttfamily 2102.11511}}].

\bibitem{Handley_dimensionality_2019}
W.~{Handley} and P.~{Lemos}, \emph{{Quantifying dimensionality: Bayesian cosmological model complexities}}, \href{https://doi.org/10.1103/PhysRevD.100.023512}{\emph{\prd} {\bfseries 100} (2019) 023512} [\href{https://arxiv.org/abs/1903.06682}{{\ttfamily 1903.06682}}].

\bibitem{2002PhRvD..66j3511L}
A.~{Lewis} and S.~{Bridle}, \emph{{Cosmological parameters from CMB and other data: A Monte Carlo approach}}, \href{https://doi.org/10.1103/PhysRevD.66.103511}{\emph{\prd} {\bfseries 66} (2002) 103511} [\href{https://arxiv.org/abs/astro-ph/0205436}{{\ttfamily astro-ph/0205436}}].

\bibitem{Handley2015PolychordI}
W.J.~{Handley}, M.P.~{Hobson} and A.N.~{Lasenby}, \emph{{polychord: nested sampling for cosmology.}}, \href{https://doi.org/10.1093/mnrasl/slv047}{\emph{\mnras} {\bfseries 450} (2015) L61} [\href{https://arxiv.org/abs/1502.01856}{{\ttfamily 1502.01856}}].

\bibitem{Handley2015PolychordII}
W.J.~{Handley}, M.P.~{Hobson} and A.N.~{Lasenby}, \emph{{POLYCHORD: next-generation nested sampling}}, \href{https://doi.org/10.1093/mnras/stv1911}{\emph{\mnras} {\bfseries 453} (2015) 4384} [\href{https://arxiv.org/abs/1506.00171}{{\ttfamily 1506.00171}}].

\bibitem{Lemos2021TensionMetrics}
P.~{Lemos}, M.~{Raveri}, A.~{Campos}, Y.~{Park}, C.~{Chang}, N.~{Weaverdyck} et~al., \emph{{Assessing tension metrics with dark energy survey and Planck data}}, \href{https://doi.org/10.1093/mnras/stab1670}{\emph{\mnras} {\bfseries 505} (2021) 6179} [\href{https://arxiv.org/abs/2012.09554}{{\ttfamily 2012.09554}}].

\bibitem{2022arXiv220711457B}
H.~{Bevins}, W.~{Handley}, P.~{Lemos}, P.~{Sims}, E.~{de Lera Acedo} and A.~{Fialkov}, \emph{{Marginal Bayesian Statistics Using Masked Autoregressive Flows and Kernel Density Estimators with Examples in Cosmology}}, \href{https://doi.org/10.3390/psf2022005001}{\emph{Physical Sciences Forum} {\bfseries 5} (2022) 1} [\href{https://arxiv.org/abs/2207.11457}{{\ttfamily 2207.11457}}].

\bibitem{Marshall2006}
P.~{Marshall}, N.~{Rajguru} and A.~{Slosar}, \emph{{Bayesian evidence as a tool for comparing datasets}}, \href{https://doi.org/10.1103/PhysRevD.73.067302}{\emph{\prd} {\bfseries 73} (2006) 067302} [\href{https://arxiv.org/abs/astro-ph/0412535}{{\ttfamily astro-ph/0412535}}].

\bibitem{2016JOSS....1...24F}
D.~{Foreman-Mackey}, \emph{{corner.py: Scatterplot matrices in Python}}, \href{https://doi.org/10.21105/joss.00024}{\emph{The Journal of Open Source Software} {\bfseries 1} (2016) 24}.

\bibitem{2023arXiv231008490O}
A.N.~{Ormondroyd}, W.J.~{Handley}, M.P.~{Hobson} and A.N.~{Lasenby}, \emph{{Tilting the scales: weighing prior dependency and global tensions of CMB lensing}}, \href{https://doi.org/10.48550/arXiv.2310.08490}{\emph{arXiv e-prints} (2023) arXiv:2310.08490} [\href{https://arxiv.org/abs/2310.08490}{{\ttfamily 2310.08490}}].

\bibitem{Planck2013params}
{Planck Collaboration}, P.A.R.~{Ade}, N.~{Aghanim}, C.~{Armitage-Caplan}, M.~{Arnaud}, M.~{Ashdown} et~al., \emph{{Planck 2013 results. XVI. Cosmological parameters}}, \href{https://doi.org/10.1051/0004-6361/201321591}{\emph{\aanda} {\bfseries 571} (2014) A16} [\href{https://arxiv.org/abs/1303.5076}{{\ttfamily 1303.5076}}].

\bibitem{cobayaascl}
J.~{Torrado} and A.~{Lewis}, ``{Cobaya: Bayesian analysis in cosmology}.'' Astrophysics Source Code Library, record ascl:1910.019, Oct., 2019.

\bibitem{Torrado2021Cobaya}
J.~{Torrado} and A.~{Lewis}, \emph{{Cobaya: code for Bayesian analysis of hierarchical physical models}}, \href{https://doi.org/10.1088/1475-7516/2021/05/057}{\emph{\jcap} {\bfseries 2021} (2021) 057} [\href{https://arxiv.org/abs/2005.05290}{{\ttfamily 2005.05290}}].

\bibitem{Lewis:1999bs}
A.~{Lewis}, A.~{Challinor} and A.~{Lasenby}, \emph{{Efficient Computation of Cosmic Microwave Background Anisotropies in Closed Friedmann-Robertson-Walker Models}}, \href{https://doi.org/10.1086/309179}{\emph{\apj} {\bfseries 538} (2000) 473} [\href{https://arxiv.org/abs/astro-ph/9911177}{{\ttfamily astro-ph/9911177}}].

\bibitem{Blumenthal1984}
G.R.~{Blumenthal}, S.M.~{Faber}, J.R.~{Primack} and M.J.~{Rees}, \emph{{Formation of galaxies and large-scale structure with cold dark matter.}}, \href{https://doi.org/10.1038/311517a0}{\emph{\nat} {\bfseries 311} (1984) 517}.

\bibitem{Riess1998}
A.G.~{Riess}, A.V.~{Filippenko}, P.~{Challis}, A.~{Clocchiatti}, A.~{Diercks}, P.M.~{Garnavich} et~al., \emph{{Observational Evidence from Supernovae for an Accelerating Universe and a Cosmological Constant}}, \href{https://doi.org/10.1086/300499}{\emph{\aj} {\bfseries 116} (1998) 1009} [\href{https://arxiv.org/abs/astro-ph/9805201}{{\ttfamily astro-ph/9805201}}].

\bibitem{Perlmutter1999}
S.~{Perlmutter}, G.~{Aldering}, G.~{Goldhaber}, R.A.~{Knop}, P.~{Nugent}, P.G.~{Castro} et~al., \emph{{Measurements of {\ensuremath{\Omega}} and {\ensuremath{\Lambda}} from 42 High-Redshift Supernovae}}, \href{https://doi.org/10.1086/307221}{\emph{\apj} {\bfseries 517} (1999) 565} [\href{https://arxiv.org/abs/astro-ph/9812133}{{\ttfamily astro-ph/9812133}}].

\bibitem{Turner1997}
M.S.~{Turner} and M.~{White}, \emph{{CDM models with a smooth component}}, \href{https://doi.org/10.1103/PhysRevD.56.R4439}{\emph{\prd} {\bfseries 56} (1997) R4439} [\href{https://arxiv.org/abs/astro-ph/9701138}{{\ttfamily astro-ph/9701138}}].

\bibitem{Chevallier2001}
M.~{Chevallier} and D.~{Polarski}, \emph{{Accelerating Universes with Scaling Dark Matter}}, \href{https://doi.org/10.1142/S0218271801000822}{\emph{International Journal of Modern Physics D} {\bfseries 10} (2001) 213} [\href{https://arxiv.org/abs/gr-qc/0009008}{{\ttfamily gr-qc/0009008}}].

\bibitem{Linder2003}
E.V.~{Linder}, \emph{{Exploring the Expansion History of the Universe}}, \href{https://doi.org/10.1103/PhysRevLett.90.091301}{\emph{\prl} {\bfseries 90} (2003) 091301} [\href{https://arxiv.org/abs/astro-ph/0208512}{{\ttfamily astro-ph/0208512}}].

\bibitem{Mangano2005}
G.~{Mangano}, G.~{Miele}, S.~{Pastor}, T.~{Pinto}, O.~{Pisanti} and P.D.~{Serpico}, \emph{{Relic neutrino decoupling including flavour oscillations}}, \href{https://doi.org/10.1016/j.nuclphysb.2005.09.041}{\emph{Nuclear Physics B} {\bfseries 729} (2005) 221} [\href{https://arxiv.org/abs/hep-ph/0506164}{{\ttfamily hep-ph/0506164}}].

\bibitem{Kosowsky1995}
A.~{Kosowsky} and M.S.~{Turner}, \emph{{CBR anisotropy and the running of the scalar spectral index}}, \href{https://doi.org/10.1103/PhysRevD.52.R1739}{\emph{\prd} {\bfseries 52} (1995) R1739} [\href{https://arxiv.org/abs/astro-ph/9504071}{{\ttfamily astro-ph/9504071}}].

\bibitem{Guth1981}
A.H.~{Guth}, \emph{{Inflationary universe: A possible solution to the horizon and flatness problems}}, \href{https://doi.org/10.1103/PhysRevD.23.347}{\emph{\prd} {\bfseries 23} (1981) 347}.

\bibitem{Starobinsky1980}
A.A.~{Starobinsky}, \emph{{A new type of isotropic cosmological models without singularity}}, \href{https://doi.org/10.1016/0370-2693(80)90670-X}{\emph{Physics Letters B} {\bfseries 91} (1980) 99}.

\bibitem{Starobinsky1979}
A.A.~{Starobinskii}, \emph{{Spectrum of relict gravitational radiation and the early state of the universe}}, {\emph{ZhETF Pisma Redaktsiiu} {\bfseries 30} (1979) 719}.

\bibitem{Skilling2004}
J.~Skilling, \emph{Nested sampling}, \href{https://doi.org/10.1063/1.1835238}{\emph{AIP Conference Proceedings} {\bfseries 735} (2004) 395}.

\bibitem{Piras2024harmonic}
D.~{Piras}, A.~{Polanska}, A.S.~{Mancini}, M.A.~{Price} and J.D.~{McEwen}, \emph{{The future of cosmological likelihood-based inference: accelerated high-dimensional parameter estimation and model comparison}}, \href{https://doi.org/10.33232/001c.123368}{\emph{The Open Journal of Astrophysics} {\bfseries 7} (2024) 73} [\href{https://arxiv.org/abs/2405.12965}{{\ttfamily 2405.12965}}].

\bibitem{Heavens2017MCEvidence}
A.~{Heavens}, Y.~{Fantaye}, A.~{Mootoovaloo}, H.~{Eggers}, Z.~{Hosenie}, S.~{Kroon} et~al., \emph{{Marginal Likelihoods from Monte Carlo Markov Chains}}, \href{https://doi.org/10.48550/arXiv.1704.03472}{\emph{arXiv e-prints} (2017) arXiv:1704.03472} [\href{https://arxiv.org/abs/1704.03472}{{\ttfamily 1704.03472}}].

\bibitem{Buchner2023}
J.~{Buchner}, \emph{{Nested Sampling Methods}}, \href{https://doi.org/10.1214/23-SS144}{\emph{Statistics Surveys} {\bfseries 17} (2023) 169} [\href{https://arxiv.org/abs/2101.09675}{{\ttfamily 2101.09675}}].

\bibitem{Ashton2022NRvMP}
G.~{Ashton}, N.~{Bernstein}, J.~{Buchner}, X.~{Chen}, G.~{Cs{\'a}nyi}, A.~{Fowlie} et~al., \emph{{Nested sampling for physical scientists}}, \href{https://doi.org/10.1038/s43586-022-00121-x}{\emph{Nature Reviews Methods Primers} {\bfseries 2} (2022) 39} [\href{https://arxiv.org/abs/2205.15570}{{\ttfamily 2205.15570}}].

\bibitem{Hu2023aeons}
Z.~{Hu}, A.~{Baryshnikov} and W.~{Handley}, \emph{{AEONS: approximating the end of nested sampling}}, \href{https://doi.org/10.1093/mnras/stae1754}{\emph{\mnras} {\bfseries 532} (2024) 4035} [\href{https://arxiv.org/abs/2312.00294}{{\ttfamily 2312.00294}}].

\bibitem{Higson2019dns}
E.~{Higson}, W.~{Handley}, M.~{Hobson} and A.~{Lasenby}, \emph{{Dynamic nested sampling: an improved algorithm for parameter estimation and evidence calculation}}, \href{https://doi.org/10.1007/s11222-018-9844-0}{\emph{Statistics and Computing} {\bfseries 29} (2019) 891} [\href{https://arxiv.org/abs/1704.03459}{{\ttfamily 1704.03459}}].

\bibitem{Planck2020likelihoods}
{Planck Collaboration}, N.~{Aghanim}, Y.~{Akrami}, M.~{Ashdown}, J.~{Aumont}, C.~{Baccigalupi} et~al., \emph{{Planck 2018 results. V. CMB power spectra and likelihoods}}, \href{https://doi.org/10.1051/0004-6361/201936386}{\emph{\aanda} {\bfseries 641} (2020) A5} [\href{https://arxiv.org/abs/1907.12875}{{\ttfamily 1907.12875}}].

\bibitem{Planck2020lensing}
{Planck Collaboration}, N.~{Aghanim}, Y.~{Akrami}, M.~{Ashdown}, J.~{Aumont}, C.~{Baccigalupi} et~al., \emph{{Planck 2018 results. VIII. Gravitational lensing}}, \href{https://doi.org/10.1051/0004-6361/201833886}{\emph{\aanda} {\bfseries 641} (2020) A8} [\href{https://arxiv.org/abs/1807.06210}{{\ttfamily 1807.06210}}].

\bibitem{CamSpec2020}
G.~{Efstathiou} and S.~{Gratton}, \emph{{A Detailed Description of the CAMSPEC Likelihood Pipeline and a Reanalysis of the Planck High Frequency Maps}}, \href{https://doi.org/10.21105/astro.1910.00483}{\emph{The Open Journal of Astrophysics} {\bfseries 4} (2021) 8} [\href{https://arxiv.org/abs/1910.00483}{{\ttfamily 1910.00483}}].

\bibitem{BICEP2021}
P.A.R.~{Ade}, Z.~{Ahmed}, M.~{Amiri}, D.~{Barkats}, R.B.~{Thakur}, C.A.~{Bischoff} et~al., \emph{{Improved Constraints on Primordial Gravitational Waves using Planck, WMAP, and BICEP/Keck Observations through the 2018 Observing Season}}, \href{https://doi.org/10.1103/PhysRevLett.127.151301}{\emph{\prl} {\bfseries 127} (2021) 151301} [\href{https://arxiv.org/abs/2110.00483}{{\ttfamily 2110.00483}}].

\bibitem{2012MNRAS.423.3430B}
F.~{Beutler}, C.~{Blake}, M.~{Colless}, D.H.~{Jones}, L.~{Staveley-Smith}, G.B.~{Poole} et~al., \emph{{The 6dF Galaxy Survey: z{\ensuremath{\approx}} 0 measurements of the growth rate and {\ensuremath{\sigma}}$_{8}$}}, \href{https://doi.org/10.1111/j.1365-2966.2012.21136.x}{\emph{\mnras} {\bfseries 423} (2012) 3430} [\href{https://arxiv.org/abs/1204.4725}{{\ttfamily 1204.4725}}].

\bibitem{Ross2015}
A.J.~{Ross}, L.~{Samushia}, C.~{Howlett}, W.J.~{Percival}, A.~{Burden} and M.~{Manera}, \emph{{The clustering of the SDSS DR7 main Galaxy sample - I. A 4 per cent distance measure at z = 0.15}}, \href{https://doi.org/10.1093/mnras/stv154}{\emph{\mnras} {\bfseries 449} (2015) 835} [\href{https://arxiv.org/abs/1409.3242}{{\ttfamily 1409.3242}}].

\bibitem{Alam2021}
S.~{Alam}, M.~{Aubert}, S.~{Avila}, C.~{Balland}, J.E.~{Bautista}, M.A.~{Bershady} et~al., \emph{{Completed SDSS-IV extended Baryon Oscillation Spectroscopic Survey: Cosmological implications from two decades of spectroscopic surveys at the Apache Point Observatory}}, \href{https://doi.org/10.1103/PhysRevD.103.083533}{\emph{\prd} {\bfseries 103} (2021) 083533} [\href{https://arxiv.org/abs/2007.08991}{{\ttfamily 2007.08991}}].

\bibitem{Riess2021}
A.G.~{Riess}, S.~{Casertano}, W.~{Yuan}, J.B.~{Bowers}, L.~{Macri}, J.C.~{Zinn} et~al., \emph{{Cosmic Distances Calibrated to 1\% Precision with Gaia EDR3 Parallaxes and Hubble Space Telescope Photometry of 75 Milky Way Cepheids Confirm Tension with {\ensuremath{\Lambda}}CDM}}, \href{https://doi.org/10.3847/2041-8213/abdbaf}{\emph{\apjl} {\bfseries 908} (2021) L6} [\href{https://arxiv.org/abs/2012.08534}{{\ttfamily 2012.08534}}].

\bibitem{Scolnic2018}
D.M.~{Scolnic}, D.O.~{Jones}, A.~{Rest}, Y.C.~{Pan}, R.~{Chornock}, R.J.~{Foley} et~al., \emph{{The Complete Light-curve Sample of Spectroscopically Confirmed SNe Ia from Pan-STARRS1 and Cosmological Constraints from the Combined Pantheon Sample}}, \href{https://doi.org/10.3847/1538-4357/aab9bb}{\emph{\apj} {\bfseries 859} (2018) 101} [\href{https://arxiv.org/abs/1710.00845}{{\ttfamily 1710.00845}}].

\bibitem{Abbott2018}
T.M.C.~{Abbott}, F.B.~{Abdalla}, A.~{Alarcon}, J.~{Aleksi{\'c}}, S.~{Allam}, S.~{Allen} et~al., \emph{{Dark Energy Survey year 1 results: Cosmological constraints from galaxy clustering and weak lensing}}, \href{https://doi.org/10.1103/PhysRevD.98.043526}{\emph{\prd} {\bfseries 98} (2018) 043526} [\href{https://arxiv.org/abs/1708.01530}{{\ttfamily 1708.01530}}].

\bibitem{Peebles1970}
P.J.E.~{Peebles} and J.T.~{Yu}, \emph{{Primeval Adiabatic Perturbation in an Expanding Universe}}, \href{https://doi.org/10.1086/150713}{\emph{\apj} {\bfseries 162} (1970) 815}.

\bibitem{HuWhite1997}
W.~{Hu} and M.~{White}, \emph{{A CMB polarization primer}}, \href{https://doi.org/10.1016/S1384-1076(97)00022-5}{\emph{\na} {\bfseries 2} (1997) 323} [\href{https://arxiv.org/abs/astro-ph/9706147}{{\ttfamily astro-ph/9706147}}].

\bibitem{HuOkamoto2002}
W.~{Hu} and T.~{Okamoto}, \emph{{Mass Reconstruction with Cosmic Microwave Background Polarization}}, \href{https://doi.org/10.1086/341110}{\emph{\apj} {\bfseries 574} (2002) 566} [\href{https://arxiv.org/abs/astro-ph/0111606}{{\ttfamily astro-ph/0111606}}].

\bibitem{EisensteinHu1998}
D.J.~{Eisenstein} and W.~{Hu}, \emph{{Baryonic Features in the Matter Transfer Function}}, \href{https://doi.org/10.1086/305424}{\emph{\apj} {\bfseries 496} (1998) 605} [\href{https://arxiv.org/abs/astro-ph/9709112}{{\ttfamily astro-ph/9709112}}].

\bibitem{Eisenstein2005}
D.J.~{Eisenstein}, I.~{Zehavi}, D.W.~{Hogg}, R.~{Scoccimarro}, M.R.~{Blanton}, R.C.~{Nichol} et~al., \emph{{Detection of the Baryon Acoustic Peak in the Large-Scale Correlation Function of SDSS Luminous Red Galaxies}}, \href{https://doi.org/10.1086/466512}{\emph{\apj} {\bfseries 633} (2005) 560} [\href{https://arxiv.org/abs/astro-ph/0501171}{{\ttfamily astro-ph/0501171}}].

\bibitem{Cole2005}
S.~{Cole}, W.J.~{Percival}, J.A.~{Peacock}, P.~{Norberg}, C.M.~{Baugh}, C.S.~{Frenk} et~al., \emph{{The 2dF Galaxy Redshift Survey: power-spectrum analysis of the final data set and cosmological implications}}, \href{https://doi.org/10.1111/j.1365-2966.2005.09318.x}{\emph{\mnras} {\bfseries 362} (2005) 505} [\href{https://arxiv.org/abs/astro-ph/0501174}{{\ttfamily astro-ph/0501174}}].

\bibitem{Kilbinger2015}
M.~{Kilbinger}, \emph{{Cosmology with cosmic shear observations: a review}}, \href{https://doi.org/10.1088/0034-4885/78/8/086901}{\emph{Reports on Progress in Physics} {\bfseries 78} (2015) 086901} [\href{https://arxiv.org/abs/1411.0115}{{\ttfamily 1411.0115}}].

\bibitem{Howlett:2012mh}
C.~{Howlett}, A.~{Lewis}, A.~{Hall} and A.~{Challinor}, \emph{{CMB power spectrum parameter degeneracies in the era of precision cosmology}}, \href{https://doi.org/10.1088/1475-7516/2012/04/027}{\emph{\jcap} {\bfseries 2012} (2012) 027} [\href{https://arxiv.org/abs/1201.3654}{{\ttfamily 1201.3654}}].

\bibitem{Mead_2016}
A.J.~{Mead}, C.~{Heymans}, L.~{Lombriser}, J.A.~{Peacock}, O.I.~{Steele} and H.A.~{Winther}, \emph{{Accurate halo-model matter power spectra with dark energy, massive neutrinos and modified gravitational forces}}, \href{https://doi.org/10.1093/mnras/stw681}{\emph{\mnras} {\bfseries 459} (2016) 1468} [\href{https://arxiv.org/abs/1602.02154}{{\ttfamily 1602.02154}}].

\bibitem{Handley2019anesthetic}
W.~{Handley}, \emph{{anesthetic: nested sampling visualisation}}, \href{https://doi.org/10.21105/joss.01414}{\emph{The Journal of Open Source Software} {\bfseries 4} (2019) 1414} [\href{https://arxiv.org/abs/1905.04768}{{\ttfamily 1905.04768}}].

\end{thebibliography}\endgroup

\end{document}